\documentclass[a4paper,11pt,preprintnumbers,groupedaddress,nofootinbib,amssymb,eqsecnum,notitlepage,superscriptaddress]{article}
\pdfoutput=1 

\usepackage{jcappub} 

\usepackage[T1]{fontenc} 

\usepackage{here}
\usepackage{subcaption} 
\usepackage{amsmath,amsthm,amssymb,mathtools}
\usepackage{bm}
\usepackage{color}
\usepackage[normalem]{ulem}
\usepackage{ulem}
\usepackage{comment}
\usepackage{physics}
\usepackage{tikz}

\usepackage{amsfonts}
\usepackage{dcolumn}
\allowdisplaybreaks[1]
\usepackage{stackengine}

\begin{document}

\newcommand{\newc}{\newcommand}
\newc{\be}{\begin{equation}}
\newc{\ee}{\end{equation}}
\newc{\ba}{\begin{eqnarray}}
\newc{\ea}{\end{eqnarray}}
\newc{\D}{\partial}
\newc{\rH}{{\rm H}}
\newc{\rd}{{\rm d}}
\newcommand{\Mpl}{M_{\rm Pl}}
\newcommand{\rBH}{r_{s}}
\newcommand{\rc}{r_{c}}
\newcommand{\rh}{r_{h}}
\newcommand{\Xh}{X_{h}}
\newcommand{\hr}{\hat{r}}
\newcommand{\ma}[1]{\textcolor{magenta}{#1}}
\newcommand{\cy}[1]{\textcolor{cyan}{#1}}
\newcommand{\mm}[1]{\textcolor{red}{#1}}
\newcommand{\nn}{\nonumber}

\begin{flushright}
WUCG-25-05 YITP-25-63 \\
\end{flushright}

\title{Effective Field Theory of Chiral Gravitational Waves}

\author[a]{Katsuki Aoki,}
\author[b,c]{Tomohiro Fujita,}
\author[d]{Ryodai Kawaguchi,}
\author[d]{and Kazuki Yanagihara}

\affiliation[a]{Center for Gravitational Physics and Quantum Information,
Yukawa Institute for Theoretical Physics, Kyoto University, Kyoto 606-8502, Japan}
\affiliation[b]{Department of Physics, Ochanomizu University, 2-2-1 Otsuka, Tokyo 112-8610, Japan}
\affiliation[c]{Kavli Institute for the Physics and Mathematics of the Universe (WPI),
The University of Tokyo Institutes for Advanced Study, The University of Tokyo, Chiba 277-8583, Japan}
\affiliation[d]{Department of Physics, Waseda University, 3-4-1 Okubo, Shinjuku, Tokyo 169-8555, Japan} 

\emailAdd{katsuki.aoki@yukawa.kyoto-u.ac.jp}
\emailAdd{fujita.tomohiro@ocha.ac.jp}
\emailAdd{ryodai0602@fuji.waseda.jp}
\emailAdd{kazuki-y@ruri.waseda.jp}

\abstract{When a (non-)Abelian gauge field acquires an isotropic background configuration during inflation, strong gravitational waves (GWs) with parity-violating polarization, known as chiral GWs, can be produced in addition to the intrinsic unpolarized GWs. However, previous studies have analyzed individual models, leaving the generality of this phenomenon unclear.
To perform a model-independent analysis, we construct an effective field theory (EFT) of chiral GWs by extending the EFT of inflation and incorporating gauge fields.
The resulting action unifies inflationary models with a $SU(2)$ gauge field, such as chromo-natural inflation and gauge-flation, and ones with a triplet of $U(1)$ gauge fields, systematically encompassing all possible GW production mechanisms consistent with the symmetry breaking induced by the gauge field background. We find that chiral GWs are generically and inevitably produced, provided that the effective energy density of the background gauge field is positive and the gauge kinetic function is not fine-tuned to a specific time dependence.
This EFT offers a useful foundation for future phenomenological studies as well as for deepening our theoretical understanding of chiral GWs.}

\maketitle
\flushbottom

\section{Introduction}
\label{Introduction}
Cosmic inflation~\cite{Starobinsky:1980te,Sato:1980yn,Kazanas:1980tx,Guth:1980zm,Linde:1981mu,Albrecht:1982wi} is now widely accepted as the leading paradigm for describing the early universe. It not only resolves the flatness and horizon problems inherent in the standard Big Bang cosmology, but also provides a natural origin for the primordial curvature perturbations that seeded the large-scale structure of the universe. The success of this framework is supported by high-precision observations of the cosmic microwave background (CMB), which have confirmed key predictions of inflation, such as the near scale-invariance and Gaussianity of the scalar power spectrum~\cite{Planck:2018vyg,Planck:2018jri}. Nevertheless, the physical mechanism behind inflation has yet to be identified. In particular, the identity of the field driving inflation, its dynamics, and its relation to fundamental physics beyond the Standard Model are still unknown. Understanding the nature of inflation thus remains a central open question in cosmology.

A promising route toward resolving this problem lies in the study of tensor perturbations, namely primordial gravitational waves (GWs), generated during inflation. These tensor modes provide a complementary observational probe to the scalar fluctuations, and are sensitive to different aspects of the inflationary dynamics~\cite{Seljak:1996gy,Kamionkowski:1996zd}. In the simplest single-field slow-roll models, GWs originate from quantum vacuum fluctuations and are predicted to be statistically isotropic, Gaussian, and unpolarized, with their amplitude directly tied to the energy scale of inflation~\cite{Starobinsky:1979ty, Mukhanov:1981xt,Bardeen:1983qw}. However, more general models can lead to richer phenomenology, including larger non-Gaussianity or signals with circular polarization~\cite{Lue:1998mq,Barnaby:2010vf,Sorbo:2011rz,Linde:2012bt, Shiraishi:2013kxa}. Future CMB B-mode polarization observations, such as those from LiteBIRD~\cite{Matsumura:2013aja} and the Simons Observatory~\cite{SimonsObservatory:2018koc} are expected to reach the sensitivity required to detect primordial GWs.
Space-based gravitational wave detectors, such as LISA~\cite{LISA:2017pwj} and DECIGO~\cite{Kawamura:2020pcg}, as well as pulsar timing arrays~\cite{NANOGrav:2023hvm}, also have the potential to detect primordial GWs.
Importantly, these experiments will also probe their circular polarization, potentially revealing new sources beyond the vacuum~\cite{Gluscevic:2010vv, Smith:2016jqs, Thorne:2017jft,Domcke:2019zls}. The detection of such parity-violating features would offer a unique window into inflationary dynamics and the symmetries of the early universe, and could significantly narrow down the viable space of inflationary models.

Among the theoretical mechanisms capable of producing distinctive gravitational wave signatures, models involving non-Abelian gauge fields have recently garnered considerable attention. It has been shown that if $SU(2)$ gauge fields acquire a non-zero, isotropic background configuration during inflation~\cite{Maleknejad:2011jw,Adshead:2012kp}, it can source gravitational waves with both large amplitudes and strong parity-violating polarization—so-called chiral GWs~\cite{Dimastrogiovanni:2012ew,Adshead:2013nka}.  
Such models occupy a prominent position among inflationary scenarios that are capable of generating GWs detectable in the near future~\cite{LiteBIRD:2023zmo}. Furthermore, gauge fields are ubiquitous in high-energy physics and the mechanism responsible for generating chiral GWs work in more general $SU(N\ge2)$ gauge theories, suggesting a broad class of realizations~\cite{Fujita:2021eue, Fujita:2022fff}. Despite these attractive theoretical and observational features, most previous studies have focused on specific model implementations, such as gauge-flation, chromo-natural inflation and their relatives~\cite{Namba:2013kia,Obata:2016tmo,Dimastrogiovanni:2016fuu,Adshead:2016omu,Caldwell:2017chz,Adshead:2017hnc,Domcke:2018rvv,Fujita:2018ndp,Lozanov:2018kpk,Watanabe:2020ctz,Holland:2020jdh,Ishiwata:2021yne,Dimastrogiovanni:2024xvc}. As a result, the underlying mechanism for chiral GW generation and its generality have remained poorly understood.

To move beyond model-specific analyses and toward a unified understanding of the origin and generality of chiral GWs, it is desirable to adopt a framework that captures the essential physics in a systematic and model-independent manner. The effective field theory (EFT) of inflation~\cite{Creminelli:2006xe, Cheung:2007st} provides such a framework by focusing on symmetry breaking: it characterizes fluctuations around an inflationary background in terms of the spontaneous breaking of time diffeomorphism invariance by an expectation value of a clock field. This approach has been successfully applied to a wide range of inflationary scenarios, including multi-field inflation~\cite{Senatore:2010wk,Shiu:2011qw,Noumi:2012vr,Pinol:2024arz}, anisotropic inflation~\cite{Abolhasani:2015cve,Rostami:2017wiy,Gong:2019hwj}, and open quantum systems~\cite{LopezNacir:2011kk,Hongo:2018ant,Salcedo:2024smn}. EFTs of cosmic fluctuations can be even formulated without a clock field, such as the solid inflation~\cite{Endlich:2012pz, Aoki:2022ipw} and a cosmic acceleration driven by a timelike vector field~\cite{Aoki:2021wew}. The difference in the field contents is translated as the difference in the symmetry-breaking patterns, allowing us to extract their universal features. In this work, we extend the EFT framework to incorporate gauge fields with isotropic background configurations. By identifying the symmetry-breaking pattern induced by the gauge fields and constructing all operators consistent with the residual symmetries, we derive a general EFT action and investigate the tensor perturbations. This formulation enables a transparent identification of the ingredients responsible for chiral GW production, and provides a systematic tool for analyzing the consequences of symmetry breaking in the early universe. Moreover, it offers a useful foundation for future phenomenological studies aiming to connect theoretical predictions with observational data.

This paper is organized as follows.
In Sec.~\ref{Sec_EFT_of_I_IGF}, we identify the symmetry-breaking pattern induced by the background gauge field configuration and construct building blocks that respect the residual symmetry.
In Sec.~\ref{Sec_EFT_action}, we formulate the EFT action based on these building blocks. We also clarify the mapping between the EFT formulation and the conventional covariant description, and illustrate it with several examples.
In Sec.~\ref{Sec_EFT_of_CGW}, we analyze tensor perturbations and determine the conditions under which chiral gravitational waves are generated. We find that, except for two exceptional cases, chiral primordial GWs are inevitably produced in the presence of the gauge field background.
Section~\ref{Conclusion} is devoted to conclusions.
Throughout this paper, Latin letters, $a,b,c,\cdots$, are used as gauge indices and Greek letters, $\mu,\nu,\rho,\cdots$, as spacetime indices.
The symmetric and antisymmetric parts of tensor $A_{\mu\nu}$ are denoted by $A_{(\mu\nu)}=\frac{1}{2}(A_{\mu\nu}+A_{\nu\mu})$ and $A_{[\mu\nu]}=\frac{1}{2}(A_{\mu\nu}-A_{\nu\mu})$, respectively.
We adopt a natural unit $c=\hbar=k_{B}=1$.

\section{Symmetry Breaking by Isotropic Gauge Fields}
\label{Sec_EFT_of_I_IGF}
We will explore a unified description of the inflationary models that generate chiral gravitational waves (GWs). For this purpose, we assume:
\begin{itemize}
\item The background spacetime is given by the flat Friedmann--Lema\^{i}tre--Robertson--Walker (FLRW) spacetime, ${\rm d} s^2=a^2(\tau)[-{\rm d} \tau^2+{\rm d} {\bm x}^2]$.
\item The time diffeomorphism invariance is spontaneously broken.
\item The spatial rotational symmetry and a (non-)abelian gauge symmetry are spontaneously broken into their diagonal subgroup.
\end{itemize}
The first two are the same as the effective field theory (EFT) of inflation~\cite{Cheung:2007st}, while the last one is assumed to generate chiral GWs at the level of leading operators of EFT. For instance, 
even without the last assumption, one can consider the gravitational Chern-Simons coupling $\phi R\tilde{R}$~\cite{Campbell:1990fu,Lue:1998mq,Jackiw:2003pm,Satoh:2008ck} and its extensions as sources of chiral GWs. However, these operators start with at least three derivatives~\cite{Creminelli:2014wna} and should be well-suppressed in the regime of validity of the EFT. We instead consider scenarios of parity violation with the help of gauge fields, which include the chromo-natural inflation~\cite{Adshead:2012kp} and the gauge-flation~\cite{Maleknejad:2011jw,Maleknejad:2011sq}. In particular, we will see that the last assumption generically leads to a parity violation of GWs in Sec.~\ref{Sec_EFT_of_CGW}.

\subsection{Time diffeomorphism breaking}
\label{Sec_EFTofI}
Let us start with a brief review of the EFT of inflation~\cite{Cheung:2007st}. The EFT of inflation is based on the spontaneous symmetry breaking of the time diffeomorphism invariance. This symmetry breaking is typically realized by a time-dependent vacuum expectation value (VEV) of a clock field, $\bar{\phi}(\tau)=\langle \phi \rangle$. This clock field may be a fundamental field or a composite operator constructed from other fields. We can choose the unitarity gauge where the clock field's fluctuation vanishes, $\delta\phi(\tau,\bm{x})\equiv \phi(\tau,\bm{x})-\bar{\phi}(\tau) =0$. The residual symmetries are then invariance under the spatial diffeomorphisms (sDiffs). 

We define a unit vector perpendicular to the uniform-$\phi$ hypersurfaces as
\ba
&&
n_\mu = -\frac{\nabla_\mu\phi}{\sqrt{-g^{\mu\nu}\nabla_\mu \phi \nabla_\nu\phi}}=-\frac{\delta^0_\mu}{\sqrt{-g^{00}}}\;,
\label{unitvector}
\ea
and the spatial metric on the hypersurfaces as
\be
h_{\mu\nu}=g_{\mu\nu}+n_\mu n_\nu\,.
\label{threemetric}
\ee
The extrinsic curvature is defined by
\be
K_{\mu\nu}=h_\mu^{~\sigma}\nabla_\sigma n_\nu, 
\label{extrinsic_curvature}
\ee
which describes how the hyperfurfaces are embedded within the 4D spacetime. Since the EFT of inflation only respects the unbroken sDiffs, we are allowed to use $n_{\mu}, h_{\mu\nu}, K_{\mu\nu}$, their derivatives and functions of time as building blocks of the EFT action.
Hence, the most general action of perturbations is given by~\cite{Cheung:2007st}\footnote{Our definitions of the EFT coefficients are different from the conventional ones in the overall factors because we are using the conformal time.}
\be
{\cal{S}}=\int{\rm d}\tau{\rm d}^3x\sqrt{-g}\left[\frac{\Mpl^2}{2}R-c(\tau)g^{00}-\Lambda(\tau)+{\cal{L}}^{(2)}\left(\delta R_{\mu\nu\rho\sigma},\delta g^{00}, \delta K_{\mu\nu}, \nabla_{\mu}, n_{\mu}, \tau\right)\right]
\label{EFTI_action}
\ee
where $c(\tau)$ and $\Lambda(\tau)$ are time-dependent EFT parameters, and
\ba
&&
\delta g^{00}\equiv g^{00}+\frac{1}{a^2}\,,
\label{deltag00}
\\
&&
\delta K_{\mu\nu}\equiv K_{\mu\nu}-\frac{{\cal{H}}}{a} h_{\mu\nu}\,,
\label{deltaK}
\\
&&
\delta R_{\mu\nu\rho\sigma}\equiv R_{\mu\nu\rho\sigma}-2\frac{{\cal{H}}^2}{a^2}h_{\mu[\rho}h_{\sigma]\nu}+\frac{{\cal{H}}'}{a^2}\left(h_{\mu\rho}n_\nu n_\sigma-h_{\mu\sigma}n_\nu n_\rho-(\mu\leftrightarrow\nu)\right)\,.
\label{deltaR}
\ea
Here, ${\cal{H}}\equiv a'/a$ is the Hubble parameter defined by the conformal time $\tau$, and a prime denotes the derivative with respect to the conformal time.
${\cal{L}}^{(2)}$ denotes terms constructed by quadratic and higher-order in perturbations. The indices are contracted by the metric $g_{\mu\nu}$, its inverse $g^{\mu\nu}$, or the Levi-Civita tensor $\epsilon_{\mu\nu\rho\sigma}$ so that ${\cal{L}}^{(2)}$ transforms as a scalar under the sDiff.

\subsection{Spatial symmetry breaking: $E$-$B$ basis}
\label{Sec_IGF_SB}
In addition to the time diffeomorphism breaking, we assume that a VEV of non-abelian gauge fields breaks the spatial and internal symmetries into their diagonal subgroup. For concreteness, we consider the $SU(2)$ gauge fields $A^a_{\mu}~(a=1,2,3)$ in a similar way to
~\cite{Adshead:2012kp, Maleknejad:2011jw,Maleknejad:2011sq} because only the part that breaks in mixture with the spatial symmetry is essential~\cite{Fujita:2021eue, Fujita:2022fff}. The background configuration of the gauge fields is supposed to be
\begin{align}
  \bar{A}^a_{\mu}(\tau)\equiv \langle A^a_{\mu} \rangle=a(\tau)Q(\tau)\delta^a_{\mu}\,,
  \label{nontrivial_BGsol}
\end{align}
where $Q\neq 0$ characterizes the background value of the gauge fields. It is known that this background is realized as an attractor solution in chromo-natural type models~\cite{Maleknejad:2013npa, Wolfson:2021fya, Murata:2022qzz}. 

Since time diffeomorphism is broken, it is convenient to introduce the “electric” and “magnetic” fields, which provide an intuitive and 3D-covariant description of the field strength,
\begin{align}
  E^a_{\mu}&\equiv n^{\rho} F^{a}_{\mu\rho}\,,\label{e_field}\\
  B^a_{\mu}&\equiv n^{\rho} \widetilde{F}^{a}_{ \mu\rho}=\frac{1}{2} \epsilon_{\mu\nu\rho\sigma} n^\nu F^{a\rho\sigma},\label{m_field}
  \end{align}
where $F_{\mu\nu}^a$ and $\widetilde{F}^{a \mu\nu}$ are the field strength of $SU(2)$ gauge fields and its dual:
\ba
&&
F_{\mu\nu}^a\equiv \partial_\mu A_\nu^a-\partial_\nu A_\mu^a-{\cal{G}}[abc]A_\mu^bA_\nu^c,\, \\
&&
\widetilde{F}_{\mu\nu}^a\equiv \frac{1}{2}\epsilon_{\mu\nu\rho\sigma}F^{a\rho \sigma}\,,
\ea
with the gauge coupling constant ${\cal{G}}$. Here, the rank-3 and rank-4 completely anti-symmetric symbols are denoted by $[abc]$ ($[123]=+1$) and $[\mu\nu\rho\sigma]$ ($[0123]=+1$), and the Levi-Civita tensor is $\epsilon_{\mu\nu\rho\sigma}=-\sqrt{-g}[\mu\nu\rho\sigma]$.
Substituting Eq.~\eqref{nontrivial_BGsol} into the above definition, we obtain background electric and magnetic fields as
\be
\bar{E}^a_\mu =a\bar{E}\delta^a_\mu\,,
\quad
\bar{B}^a_\mu=a\bar{B}\delta^a_\mu\,,
\label{nontrivial_BGsol_EB}
\ee
with 
\be
\bar{E}\equiv -\frac{1}{a}(Q'+{\cal{H}}Q),\quad\bar{B}\equiv-{\cal{G}}Q^2\,.
\label{EbarBbar}
\ee
As seen above, the gauge coupling $({\cal{G}}\neq 0)$ inherent in non-Abelian fields enables the magnetic component, as well as the electric one, to develop isotropic background values.

Note that scenarios with a triplet of $U(1)$ gauge fields with the same isotropic background~\cite{Yamamoto:2012tq, Firouzjahi:2018wlp} can be understood as the $\mathcal{G}\to 0$ limit of our case.
We also note that the magnetic (or electric) gaugid~\cite{Piazza:2017bsd} assumes a different realization of the isotropic background for a triplet of $U(1)$ gauge fields itself, $\bar{A}^a_{\mu}=[abc]\delta^b_{\mu}\delta^c_{\nu}x^{\nu}$ ($\bar{A}^a_\mu=\delta_\mu^0\delta_\nu^a x^\nu$), which results in a non-zero isotropic magnetic (electric) background while has no electric (magnetic) background.
Our EFT introduced in Sec.~\ref{Sec_EFT_action} is applicable to the gaugid background as well because our construction only requires that either $\bar{E}$ or $\bar{B}$ is non-zero. In the following, however, we only focus on the background \eqref{nontrivial_BGsol} with assuming $\bar{E}\neq 0$ for simplicity.

The background~\eqref{nontrivial_BGsol_EB} is invariant only under a simultaneous rotation of the global $SO(3)$ part of the sDiffs and the $SU(2)$ transformation, yielding a spontaneous symmetry breaking~\cite{Adshead:2012kp, Maleknejad:2011jw}. For our purpose, it is more transparent to rephrase the setup in the language of the local frame~\cite{Maleknejad:2011sq} by introducing a spatial coframe $e^A_\mu$, namely, triads that satisfy the following relations
\be
\delta_{AB} e^A_\mu e^B_\nu =h_{\mu\nu}\,,\quad
h^{\mu\nu} e^A_\mu e^B_\nu =\delta^{AB}\,,\quad
n^\mu e^{A}_{\mu}=0\,,
\label{basis_vecotors}
\ee
where the indices $A, B,\cdots$ represent the spatial components in the local Lorentz coordinates and run from $1$ to $3$. The background of the triads is chosen to be $\bar{e}^A_{\mu}=a \delta^A_{\mu}$. With this choice, the global $SO(3)$ part of sDiffs is identified with the global rotation of the triads, so we can rephrase our symmetry breaking as the breaking of the global parts of the $SU(2)_I$ internal rotation and the coframe $SO(3)_S$ spatial rotation into their diagonal part,
\begin{align}
    \text{global}:SU(2)_I \times SO(3)_S \to SO(3)_D
    \,,\label{symmetry_breaking_global}
\end{align}
where the subscripts are added to distinguish different symmetries (See Appendix.~\ref{AppendixA} for a detailed discussion).

With the global symmetry-breaking pattern and the residual symmetries clarified, we now turn to the local symmetry. By appropriately fixing the gauge, we will systematically construct the effective action for perturbations. Thanks to the triads, we have two rotational local symmetries, $SU(2)_I$ for the gauge fields and $SO(3)_S$ for the spatial rotation, in addition to sDiffs. In the presence of the background~\eqref{nontrivial_BGsol_EB}, and more specifically when $\bar{E} \neq 0$, we can fix the gauge such that the perturbations satisfy\footnote{The possibility of the gauge fixing \eqref{gauge_fixing} can be easily shown, for example, from the fact that the $SO(3)$ generators are antisymmetric tensors $\omega_{AB}=\omega_{[AB]}$ and the VEVs of $E^a_{\mu}$ and $e^A_{\mu}$ are proportional to the Kronecker delta (see Appendix.~\ref{AppendixA}). If one wants to consider the background $\bar{E}=0$ but $\bar{B}\neq 0$, one may instead consider the gauge fixing $\delta_{aA} B^a_{[\mu} e^A_{\nu]}=0$.}
\begin{align}
    \delta_{aA} E^a_{[\mu} e^A_{\nu]}=0\,.
\label{gauge_fixing}
\end{align}
This gauge fixing constrains (or effectively ``breaks'') the independent local rotations of the spatial and internal symmetry groups while preserving the simultaneous local rotations,
\begin{align}
    \text{local}: SU(2)_I \times SO(3)_S \to SO(3)_D
    \,.\label{symmetry_breaking}
\end{align}
Hence, the local $SU(2)_I$ and $SO(3)_S$ are broken in the same manner as their global counterparts~\eqref{symmetry_breaking_global}. In particular, this gauge fixing enables us to identify the local spatial indices with the internal gauge indices even at the level of perturbations because $\delta_{aA}$ remains invariant under the unbroken local $SO(3)_D$ transformation. Hence, we write the triads as $e^a_{\mu}$ in the following.

Therefore, to construct the EFT, our task is simply to find the most general action of the perturbations under the sDiffs and the $SO(3)_D$ symmetry. We define the perturbations of the electric and magnetic fields by
\be
    \delta E^a_{\mu}\equiv  E^a_{\mu}-\bar{E}e^a_{\mu}\,, \qquad \delta B^a_{\mu}\equiv B^a_{\mu}-\bar{B}e^a_{\mu}\,,
    \label{EB_perturbations}
\ee
which transform as tensors under our residual symmetries. We refer to these building blocks as the $E$-$B$ basis. Since the triads satisfy \eqref{basis_vecotors}, by contracting the above perturbations with the triads, we construct the $E$-$B$ perturbations only with the spacetime indices
\begin{align}
    \delta E_{\mu\nu} \equiv \delta E^a_{\mu}e^a_{\nu} = E^a_{\mu}e^a_{\nu}-\bar{E}h_{\mu\nu}
    \,, \quad
    \delta B_{\mu\nu} \equiv \delta B^a_{\mu}e^a_{\nu} = B^a_{\mu}e^a_{\nu}-\bar{B}h_{\mu\nu}
    \,,
\end{align}
where the repeated $a$ indices are contracted. The inverse relations are
\begin{align}
    \delta E^a_{\mu}= h^{\nu\rho} \delta E_{\mu\nu}e^a_{\rho}\,, \qquad \delta B^a_{\mu}= h^{\nu\rho} \delta B_{\mu\nu}e^a_{\rho}\,.
    \label{EB^a_mu}
\end{align}
Here, $\delta E_{\mu\nu}$ is symmetric in its indices because of the gauge condition \eqref{gauge_fixing}, whereas $\delta B_{\mu\nu}$ has no symmetry.\footnote{While $\delta E^a_{\mu}$ and $\delta B^a_{\mu}$ have $9\times 2=18$ components corresponding to the number of independent components of $F^a_{\mu\nu}$, the tensors $\delta E_{\mu\nu}$ and $\delta B_{\mu\nu}$ only have $6+9=15$ components. The missing 3 components are the gauge modes of the $SO(3)_D$ transformation. Hence, $\delta E_{\mu\nu}$ and $\delta B_{\mu\nu}$ have all information of the gauge fields $\delta E_\mu^a$ and $\delta B_\mu^a$, and thus $F^a_{\mu\nu}$ up to the gauge freedom.} They are both spatial in the sense that they are orthogonal to the normal vector, e.g. $n^{\mu}\delta E_{\mu\nu}=0$. The $SO(3)_D$ symmetry becomes manifest by the use of $\delta E_{\mu\nu}$ and $\delta B_{\mu\nu}$. The invariant Lagrangian is obtained by contracting the spacetime indices.

At this point, it is worth clarifying why the symmetry-breaking $SU(2)_I\times SO(3)_S \to SO(3)_D$ will lead to parity violation. The key point is that the background~\eqref{nontrivial_BGsol} contains a non-zero magnetic component, which is odd under parity. As a result,  the background is not invariant under the parity transformation. This should be contrasted with the case of a triplet of $U(1)$ gauge fields with the global $SU(2)$ symmetry~\cite{Yamamoto:2012tq, Firouzjahi:2018wlp}, which can be regarded as the $\mathcal{G}\to 0$ limit of our case. In the $U(1)$ case with the background \eqref{nontrivial_BGsol}, it has no magnetic field and parity is not violated through gauge field dynamics alone, since $U(1)$ lacks self-interactions. Parity violation can be induced, for instance, by a pseudoscalar field through the Chern-Simons coupling $\phi F \widetilde{F}$~\cite{Anber:2006xt,Sorbo:2011rz}.
In the $SU(2)$ gauge fields case, on the other hand, we will see that the parity of GWs is inevitably broken regardless of whether there is a pseudoscalar field under the conditions we clarify later.

Let us make a comparison with the solid inflation where the spatial symmetry is also broken~\cite{Endlich:2012pz}. The solid inflation is based on a triplet of scalar fields, $\phi^I$ ($I=1,2,3$), with the global internal rotation and translation symmetries, $\phi^I\rightarrow O^I{}_J\phi^J$ ($O^I{}_J\subset SO(3)$), and $\phi^I\rightarrow\phi^I+a^I$ ($a^I=\text{const.}$). The scalar fields are assumed to have the background configuration $\langle\phi^I\rangle=x^I$, which leads to the spontaneous symmetry breaking $SO(3)_I \times SO(3)_S \to SO(3)_D$ (the translation is also broken), similarly to \eqref{symmetry_breaking_global}. However, the residual symmetries of the models with gauge fields and the solid inflation are different: the former possesses the {\it local} $SO(3)_D$ symmetry while the latter only has the {\it global} $SO(3)_D$ symmetry. In particular, these models have different numbers of degrees of freedom in the tensor modes. The gauge field models have four tensor modes in the broken phase, where two originate from the metric and other two are from the gauge fields. On the other hand, the solid inflation only has two tensor modes from the metric, acquiring a mass due to the symmetry breaking.

\subsection{Alternative building blocks: $X$-$Y$ basis}
\label{Sec_BD_gauge_field}
In the previous subsection, we discussed that our EFT is defined by the symmetry-breaking pattern \eqref{symmetry_breaking} in addition to the time diffeomorphism breaking, and the building blocks for the EFT are the perturbations of the electric and magnetic fields \eqref{EB_perturbations}. These building blocks are useful for understanding the symmetry-breaking patterns and rewriting the action given by $F^a_{\mu\nu}$ into the EFT form. In practical calculations, however, the $E$-$B$ basis would be less convenient because it requires the use of the 9 components triads $e^a_{\mu}$, instead of the 6 components spatial metric $h_{\mu\nu}$, with the additional constraint \eqref{gauge_fixing} to eliminate the extra 3 components. In this subsection, we introduce an alternative basis to bypass the use of triads.

The new building blocks and their perturbations are defined by
\begin{align}
X_{\mu\nu}&\equiv 
\frac{1}{2}\left(E^a_\mu E^a_\nu-B^a_\mu B^a_\nu \right)\,,
\label{X_EB_exact_relation}
\\
Y_{\mu\nu}&\equiv E^a_\mu B^a_\nu\,,
\label{Y_EB_exact_relation}
\end{align}
and
\begin{align}
\delta X_{\mu\nu} &\equiv X_{\mu\nu}-\bar{X}_{\mu\nu} = X_{\mu\nu}-{
\frac{1}{2} }\left[\bar{E}^2-\bar{B}^2\right]h_{\mu\nu}\,, \label{deltaX}\\
\delta Y_{\mu\nu}&\equiv Y_{\mu\nu}-
\bar{Y}_{\mu\nu}= Y_{\mu\nu}-\bar{E}\bar{B}h_{\mu\nu}\,,
\label{deltaY}
\end{align}
which do not include the triads in their definitions. They are invariant under the $SU(2)_I$ transformation and covariant with respect to the sDiffs, so any scalar constructed by $\delta X_{\mu\nu}, \delta Y_{\mu\nu}$ and other sDiffs tensors respects the residual symmetries of the EFT. 

The question is whether the effective action spanned by $\delta X_{\mu\nu}$ and $\delta Y_{\mu\nu}$, which is called the $X$-$Y$ basis, indeed covers all the possible theory space in the broken phase. 
Having established that the EFT action can be written in the $E$-$B$ basis, our remaining task is to show the equivalence between this basis and the $X$-$Y$ basis.
We first count the number of independent components. $\delta X_{\mu\nu}$ is symmetric and $\delta X_{\mu\nu}, \delta Y_{\mu\nu}$ are spatial in the sense of $n^\mu \delta X_{\mu\nu}=n^\mu \delta Y_{\mu\nu}=n^\nu \delta Y_{\mu\nu}=0$, which implies that there are $6+9=15$ independent components. This is the same number as the independent components of $\delta E_{\mu\nu}$ and $\delta B_{\mu\nu}$. 
While this count suggests that the two bases may be equivalent, the equivalence becomes manifest once we explicitly construct the one-to-one correspondence between them. It is straightforward to express $\delta X, \delta Y$ in terms of $\delta E, \delta B$:
\begin{align}
    \delta X_{\mu\nu}&=
    \bar{E}e^a_{(\mu} \delta E^a_{\nu)} - \bar{B} e^a_{(\mu}\delta B^a_{\nu)}+ \frac{1}{2}\delta E^a_{\mu}\delta E^a_{\nu} -\frac{1}{2}\delta B^a_{\mu}\delta B^a_{\nu} 
    \nonumber 
    \label{XbyEB}\\
    &
    =  \bar{E} \delta E_{\mu\nu}-\bar{B} \delta B_{(\mu\nu)} + \frac{1}{2}\delta E_{\mu}{}^{\alpha}\delta E_{\nu\alpha} - \frac{1}{2}\delta B_{\mu}{}^{\alpha}\delta B_{\nu\alpha} 
    \,,  \\
    \delta Y_{\mu\nu}&=\bar{B} \delta E^a_{\mu}e^a_{\nu} +\bar{E} e^a_{\mu}\delta B^a_{\nu}+ \delta E^a_{\mu}\delta B^a_{\nu} 
    \nonumber \\
    &=\bar{B} \delta E_{\mu\nu} +\bar{E} \delta B_{\nu \mu}+\delta E_{\mu}{}^{\alpha}\delta B_{\nu \alpha}
    \,.
    \label{YbyEB}
\end{align}
The inverse formula is obtained by perturbatively solving these equations for $\delta E_{\mu\nu}$ and $\delta B_{\mu\nu}$.
We write
\begin{align}
\delta E_{\mu\nu}&=
\delta E^{(1)}_{\mu\nu}+\delta E^{(2)}_{\mu\nu}+\cdots,\\
\delta B_{\mu\nu}&=
\delta B^{(1)}_{\mu\nu}+\delta B^{(2)}_{\mu\nu}+\cdots\,,
\end{align}
where $(1),(2),\cdots$ denote the order of $\delta X_{\mu\nu},\delta Y_{\mu\nu}$.
The first order equations are
\begin{align}
\delta X_{\mu\nu}&={
\bar{E}\delta E^{(1)}_{\mu\nu}-\bar{B}\delta B^{(1)}_{(\mu\nu)}},\\
\delta Y_{\mu\nu}&=\bar{B}\delta E^{(1)}_{\mu\nu}+\bar{E}\delta B^{(1)}_{\nu\mu},
\end{align}
which can be solved as
\ba
&&
\delta E^{(1)}_{\mu\nu}=\frac{1}{\bar{E}^2+\bar{B}^2}(\bar{E}\delta X_{\mu\nu}+\bar{B}\delta Y_{(\mu\nu)})\,, 
\label{X_EB_first}
\\
&&
\delta B^{(1)}_{\mu\nu}=\frac{1}{\bar{E}^2+\bar{B}^2}(-\bar{B}\delta X_{\mu\nu}+\bar{E}\delta Y_{(\mu\nu)})-\frac{1}{\bar{E}}\delta Y_{[\mu\nu]} 
\label{Y_EB_first}
\,,
\ea
under the gauge condition $\delta E_{[\mu\nu]}=0$ and $\bar{E}\neq 0$. For the second order, we have
\begin{align}
0&=
\frac{1}{2}(\delta E^{(1)}_{\mu\rho}\delta E^{(1)}_{\nu\sigma}h^{\rho\sigma}-\delta B^{(1)}_{\mu\rho}\delta B^{(1)}_{\nu\sigma}h^{\rho\sigma})+(\bar{E}\delta E^{(2)}_{(\mu\nu)}-\bar{B}\delta B^{(2)}_{(\mu\nu)})\label{iterate_second},\\
0&=\delta E^{(1)}_{\mu\rho}\delta B^{(1)}_{\nu\sigma}h^{\rho\sigma}+\bar{B}\delta E^{(2)}_{\mu\nu}+\bar{E}\delta B^{(2)}_{\nu\mu},
\end{align}
and obtain
\ba
&&
\delta E^{(2)}_{\mu\nu}=\frac{-1}{2(\bar{E}^2+\bar{B}^2)}\left[\bar{E}(\delta E^{(1)}_{\mu\rho}\delta E^{(1)}_{\nu\sigma}-\delta B^{(1)}_{\mu\rho}\delta B^{(1)}_{\nu\sigma})+2\bar{B}\delta E^{(1)}_{\rho(\mu}\delta B^{(1)}_{\nu)\sigma}\right]h^{\rho\sigma}\,,
\label{X_EB_second}
\\
&&
\delta B^{(2)}_{\mu\nu}=\frac{1}{2(\bar{E}^2+\bar{B}^2)}\left[\bar{B}(\delta E^{(1)}_{\mu\rho}\delta E^{(1)}_{\nu\sigma}-\delta B^{(1)}_{\mu\rho}\delta B^{(1)}_{\nu\sigma})-2\bar{E}\delta E^{(1)}_{\rho(\mu}\delta B^{(1)}_{\nu)\sigma}\right]h^{\rho\sigma}
\notag
\\
&&
\hspace{1.5cm}
+\frac{1}{\bar{E}}\delta E^{(1)}_{\rho[\mu}\delta B^{(1)}_{\nu]\sigma}h^{\rho\sigma}\,.\label{Y_EB_second}
\ea
By substituting \eqref{X_EB_first} and \eqref{Y_EB_first} into them, we obtain the inverse formula up to the second order in perturbations. It is cumbersome but straightforward to find higher-order solutions for $\delta E_{\mu\nu}$ and $\delta B_{\mu\nu}$ by repeating these procedures provided $\bar{E}\neq 0$.\footnote{The $X$-$Y$ basis itself can be used even for $\bar{E}=0$ if $\bar{B}\neq 0$. In this case, one can use the gauge condition $\delta_{aA}B^a_{[\mu}e^A_{\nu]}=0$ to define the new basis and show the equivalence of this to the $X$-$Y$ basis.} If necessary, $\delta E^a_{\mu}$ and $\delta B^a_{\mu}$ are obtained by applying \eqref{EB^a_mu}.

In summary, the most general action of the perturbations of the gauge fields can be written by either $\delta E^a_{\mu}$ and $\delta B^a_{\mu}$ or $\delta X_{\mu\nu}$ and $\delta Y_{\mu\nu}$. Both bases are covariant under the residual symmetries. One may wonder why the symmetry breaking is less obvious in the $X$-$Y$ basis, which is invariant under the independent $SU(2)_I$ transformation rather than the diagonal rotation $SO(3)_D$. It should be recalled that local symmetries are just redundancies of description, and a local symmetry can be hidden when we use variables that are free from redundancies. The metric is invariant under local rotations of the coframe. Hence, in the $X$-$Y$ basis, what we can see is the breaking of the global parts of sDiffs and $SU(2)_I$ only. We emphasize again that $(\delta X_{\mu\nu},\delta Y_{\mu\nu})$ covering all the theory space is justified under the existence of the non-zero VEV of the gauge fields \eqref{nontrivial_BGsol}, and the information of symmetry breaking is used here.

\section{EFT action}
\label{Sec_EFT_action}
\subsection{Leading operators}
Now, we have all the building blocks for the EFT in the presence of the isotropic gauge field background. As we have mentioned, we are interested in the leading operators in the derivative expansion. Before the explicit construction of the EFT action, let us begin with clarifying what the leading operators mean.

As is well known, the dimensional counting in a non-trivial background can be different from that in a trivial background.
Let us see some explicit examples in the flat spacetime.
The prominent example is a k-essence scalar field, where the Lagrangian is given by
\begin{align}
    \mathcal{L}=\sum_{n=1}^{\infty} \frac{a_n}{M^{4(n-1)}}(\nabla \phi)^{2n}
    \,,
    \label{k-essence}
\end{align}
with $M$ being a constant of mass dimension one.
The $n\geq 2$ operators are sub-leading around the trivial background $\bar{\phi}=0$.
On the other hand, they all can have the same importance as the $n=1$ operator if the background has a large velocity $\bar{\phi} \sim M^2 t$. 
Substituting $\phi=M^2 t+\delta\phi$ into the Lagrangian, we obtain
\ba
&&
(n=1),\quad (\nabla\phi)^2\supset(\nabla\delta\phi)^2\sim k^2\delta\phi^2 \,,
\\
&&
(n\ge2),\quad
\frac{1}{M^{4(n-1)}}(\nabla\phi)^{2n}\supset \frac{1}{M^{4(n-1)}}(\nabla\bar{\phi})^{2(n-1)}(\nabla\delta\phi)^2\sim k^2\delta\phi^2\,.
\ea
Here, only leading perturbations are shown in each operator and we assume that $\delta\phi$ fluctuates with a momentum $k$.
We can see that $n\ge 2$ operators lead to the same order as $n=1$ operator due to the non-trivial background with the large velocity $\bar{\phi}=M^2 t$.
Note that the EFT is still valid even around such a large background because higher-derivative operators are well suppressed as long as the fluctuations $\delta \phi$ have a low-momentum $k\ll M$.
For instance,
\ba
&&
\frac{1}{M^2}(\nabla^2 \phi)^2\supset\frac{1}{M^2}(\nabla^2\delta\phi)^2\sim \frac{k^4}{M^2}\delta\phi^2 \,,
\\
&&
\frac{1}{M^6}(\nabla \phi)^2 (\nabla^2 \phi)^2\supset\frac{1}{M^6}(\nabla \bar{\phi})^2 (\nabla^2 \delta\phi)^2\sim \frac{k^4}{M^2}\delta\phi^2\,,
\ea
which are suppressed by $k^2/M^2\ll1$ compared to the leading operators \eqref{k-essence}. In the following, we only focus on quadratic terms in perturbations and count the number of the highest-order derivatives acting on perturbations. Then, the discussions so far can be summarised as
\begin{align}
    \frac{1}{M^{4(n-1)}}(\nabla \phi)^{2n}\sim k^2\,, \qquad \frac{1}{M^{4n+2}}(\nabla \phi)^{2n} (\nabla^2 \phi)^{2n}\sim \frac{k^4}{M^2} \ll k^2
    \,.
\end{align}
The first one is leading, whereas the second one is sub-leading within the regime of validity of the EFT, $k\ll M$.

Let us now apply the same counting rule of derivatives to concrete examples.
Another example, especially relevant for us, is the chromo-natural inflation~\cite{Adshead:2012kp}, which relies on the Chern-Simons coupling
\begin{align}
    \frac{\lambda}{f} \phi\, F^a_{\mu\nu}\widetilde{F}^{a\mu\nu} = - \frac{\lambda}{f} \nabla_{\mu}\phi\, J^{\mu} + (\text{boundary terms})\,,
\end{align}
where $J^{\mu}$ is defined by $F^a_{\mu\nu}\widetilde{F}^{a\mu\nu}=\nabla_{\mu}J^{\mu}$, which only yields up to the first derivative of perturbations.
Therefore, if $\phi$ has a background velocity, we can count the number of derivatives as
\be
\frac{\lambda}{f} \phi\, F^a_{\mu\nu}\widetilde{F}^{a\mu\nu} \sim \frac{\lambda M^2}{f}k\,.
\ee
This operator is thus more important than the kinetic term $FF \sim k^2$ in the low-energy regime $\lambda M^2/f \gg k$.
The other example is the gauge-flation~\cite{Maleknejad:2011jw,Maleknejad:2011sq} whose gauge part of the Lagrangian is given by
\begin{align}
   {\cal{L}}= -\frac{1}{4}F^a_{\mu\nu}{F}^{a\mu\nu}+\frac{1}{16M^4}\left(F^a_{\mu\nu}\widetilde{F}^{a\mu\nu}\right)^2
    \,.
\end{align}
Similarly to the k-essence, the operator $(F^a_{\mu\nu}\widetilde{F}^{a\mu\nu})^2$ becomes leading if the gauge field has a large background $\bar{E},\bar{B}\sim M^2$, that is, $ \frac{1}{M^4}F^4 \sim k^2 $.
On the other hand, operators involving derivatives of the field strength are sub-leading as they yield higher derivatives, e.g., $\frac{1}{M^2}(\nabla F)^2 \sim \frac{k^4}{M^2}$.

Gravitational couplings can be counted similarly. The couplings $f(\phi)R$ involve only up to second derivatives, and thus can be leading operators. The gravitational Chern-Simons coupling we mentioned at the beginning of Sec.~\ref{Sec_EFT_of_I_IGF} is
\be
\frac{\phi}{f}\epsilon^{\mu\nu\rho\sigma}R_{\alpha\beta\mu\nu}R^{\alpha\beta}{}_{\rho\sigma} =-\frac{1}{f}\nabla_{\mu} \phi \, \mathcal{J}^{\mu} + (\text{boundary terms}) \sim \frac{M^2}{f \Mpl^2} k^3
\ee
where we have first performed the integration by parts and used $\bar{\phi}=M^2 t$. The Planck mass $\Mpl$ appears after canonical normalisation of metric perturbations.
Hence, the gravitational Chern-Simons coupling is third order in derivatives, and thus sub-leading. One can also discuss derivative couplings to the curvature, such as $R(\nabla \phi)^2$ and $R F^2$, which are higher derivatives and should be suppressed, in general.

All in all, the leading operators here refer to leading in the sense of derivative expansion, rather than counting mass dimensions. These operators describe low-energy/momentum dynamics of perturbations around a non-trivial background. Note that there are exceptions to the discussion thus far. It is well known, in the context of modified gravity like Horndeski theories~\cite{Horndeski:1974wa}, that certain combinations of higher derivative operators generate equations of motion up to second order in derivatives. These combinations can be counted as leading by following the rule of the number of derivatives, albeit requiring two mass scales often denoted by $\Lambda_2$ and $\Lambda_3$ with the hierarchy $\Lambda_2 \gg \Lambda_3$~\cite{Pirtskhalava:2015nla}.
For the sake of simplicity, this paper will consider the conservative case where no such hierarchy exists,\footnote{It has been shown in~\cite{Bellazzini:2020cot, Tolley:2020gtv, Caron-Huot:2020cmc} that this hierarchy obstructs UV completion (or renders a low cutoff) at least around the Minkowski background.} even though the framework of our EFT can easily extend to accommodate such combinations.

Under these rules, let us identify all leading operators and extract their universal features.
In the language of $\phi$ and $F^a_{\mu\nu}$, the most general Lagrangian takes the form
\begin{align}
    \mathcal{L}=\frac{M_{\rm Pl}^2}{2}R + {\cal{L}}_M(g_{\mu\nu},\phi,\nabla_{\mu} \phi, F^a_{\mu\nu}) + \cdots
\end{align}
where $\cdots$ are operators containing more than one derivative of fields, e.g. $\nabla_{\mu}\nabla_{\nu}\phi, \nabla_{\mu} F^a_{\nu\rho}$ and derivative couplings to the curvature, which we ignore in the following. On the other hand, we keep all non-linear terms of first derivatives of the fields because, as we have discussed, they can have the same importance in the broken phase. We define our EFT in the Einstein frame by eliminating the coupling $f(\phi)R$ through a field redefinition so that the leading gravitational part of the Lagrangian takes the form of the Einstein-Hilbert action.\footnote{
When applying the EFT to dark energy, it may be useful to define the EFT in the Jordan frame where the ordinary matter fields are kept minimally coupled to gravity~\cite{Gubitosi:2012hu}.} Using the building blocks developed in Sec.~\ref{Sec_EFT_of_I_IGF}, we can write the same Lagrangian in a different basis:
\begin{align}
     \mathcal{L} =
         \frac{M_{\rm Pl}^2}{2}R + {{\cal{L}}_M}(h_{\mu\nu},\tau ,\delta g^{00}, \delta E_{\mu\nu}, \delta B_{\mu\nu}) + \cdots
         \,,
\end{align}
and
\begin{align}
     \mathcal{L} =
     \frac{M_{\rm Pl}^2}{2}R + {\cal{L}}_M(h_{\mu\nu},\tau ,\delta g^{00}, \delta X_{\mu\nu}, \delta Y_{\mu\nu}) + \cdots
     \,,
     \label{L_XY}
\end{align}
where $\delta K_{\mu\nu}$ and $\delta R_{\mu\nu\rho\sigma}$, which appear in the action of the EFT of inflation~\eqref{EFTI_action}, are included in the ellipsis and thus neglected in the following.\footnote{ $K_{\mu\nu}$ is related to a higher derivative operator $\nabla_\mu\nabla_\nu\phi$ through the definitions of the unit vector~\eqref{unitvector} and the extrinsic curvature~\eqref{extrinsic_curvature}.} In this subsection, we adopt the $X$-$Y$ basis \eqref{L_XY}, and we will discuss how the Lagrangian is mapped into a different basis in Sec.~\ref{Sec_mapping}.

Organizing the Lagrangian up to  quadratic order in perturbations, we obtain the most general expression in the unitary gauge as
\begin{align}
    \mathcal{L}&=\frac{M_{\rm Pl}^2}{2}R-\Lambda(\tau)-c(\tau)g^{00}+\lambda_X(\tau)X+\lambda_Y(\tau)Y + {\cal{L}}^{(2)}
    \,, \label{EFT_full}
\\
 {\cal{L}}^{(2)}&= M_{gg}^4(\tau)(\delta g^{00})^2+ M_{gX}^2(\tau)\delta g^{00}\delta X+ M_{gY}^2(\tau)\delta g^{00}\delta Y
\nonumber \\
&+\frac{1}{M_{XX}^4(\tau)}\delta X^2+\frac{1}{M_{XY}^4(\tau)}\delta X\delta Y+\frac{1}{M_{YY}^4(\tau)}\delta Y^2
\nonumber \\
&
+\frac{1}{\bar{M}_{XX}^4(\tau)}\delta X_{\mu\nu}\delta X^{\mu\nu}
\nonumber
+\frac{1}{\bar{M}_{XY}^4(\tau)}\delta X_{\mu\nu}\delta Y^{\mu\nu}
+\frac{1}{\bar{M}_{YY}^4(\tau)}\delta Y_{(\mu\nu)}\delta Y^{(\mu\nu)}
\nonumber \\
&
+\frac{1}{\hat{M}_{YY}^4(\tau)}\delta Y_{[\mu\nu]}\delta Y^{[\mu\nu]}
+\cdots,
\label{EFT_F2}
\end{align}
where $X\equiv g^{\mu\nu}X_{\mu\nu}=h^{\mu\nu}X_{\mu\nu}$ and $Y, \delta X, $ and $\delta Y$ are defined analogously. Terms beyond quadratic order in perturbations are omitted and indicated by $\cdots$. Each coefficient is generically a function of the conformal time $\tau$. The operators $\lambda_X X$ and $\lambda_Y Y$ are the new tadpole contributions from the gauge field, while ${\cal{L}}^{(2)}$ starts at the quadratic order in perturbations. It is worth noting that
\ba
&&
X=-
\frac{1}{4}F_{\mu\nu}^aF^{a\mu\nu}\,,
\label{X_FF}
\\
&&
Y=-\frac{1}{4}F_{\mu\nu}^a\widetilde{F}^{a\mu\nu}\,.
\label{Y_FFtilde}
\ea
Thus, one can see that the tadpole terms are related to the standard kinetic term and the Chern-Simons coupling
\begin{align}
    \mathcal{L}&=\frac{M_{\rm Pl}^2}{2}R-\Lambda(\tau)-c(\tau)g^{00} 
    -\frac{\lambda_X(\tau)}{4}F^a_{\mu\nu}F^{a\mu\nu}-\frac{\lambda_Y(\tau)}{4}F^a_{\mu\nu}\widetilde{F}^{a\mu\nu} + {\cal{L}}^{(2)}\,.
\end{align}
However, as with the EFT of inflation, the EFT coefficients $\lambda_X$ and $\lambda_Y$ contain the information of all non-linear terms in the language of $\phi$ and $F^a_{\mu\nu}$, and in particular, $\lambda_Y$ does not necessarily originate from the Chern-Simons coupling as we will see shortly.

\subsection{Mapping between different bases}
\label{Sec_mapping}
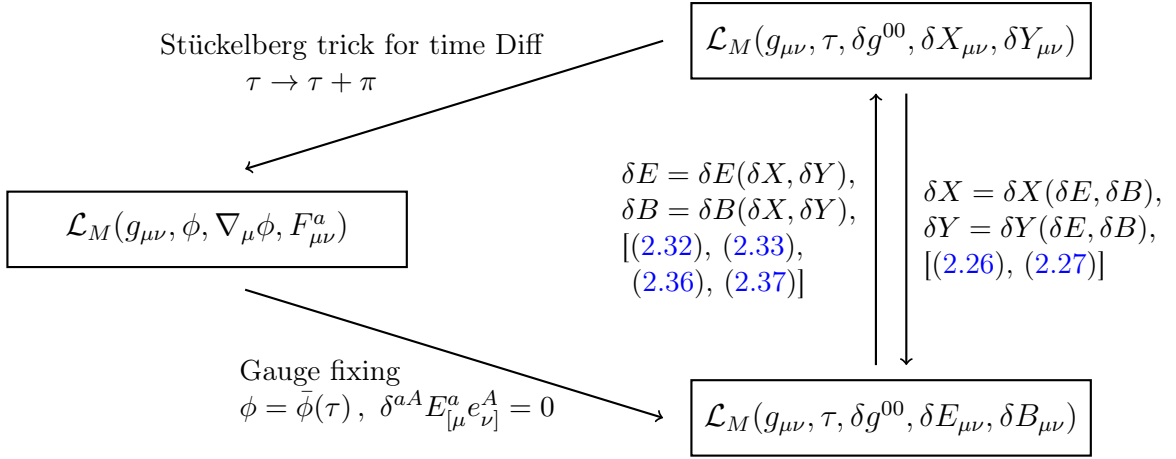
\begin{figure}[t]
    \centering
    \begin{tikzpicture}
    \node[rectangle,thick,draw=black,text width=5cm,text centered,minimum height=1cm] at (-4.5,0) {\large{${\cal{L}}_M(g_{\mu\nu},\phi,\nabla_{\mu} \phi, F^a_{\mu\nu})$}};
    \node[rectangle,thick,draw=black,text width=5cm,text centered,minimum height=1cm] at (4.5,2.5) {\large{${\cal{L}}_M(g_{\mu\nu},\tau,\delta g^{00},\delta X_{\mu\nu}, \delta Y_{\mu\nu})$}};
    \node[rectangle,thick,draw=black,text width=5cm,text centered,minimum height=1cm] at (4.5,-2.5) {\large{${\cal{L}}_M(g_{\mu\nu},\tau,\delta g^{00},\delta E_{\mu\nu}, \delta B_{\mu\nu})$}};
    \draw[->,thick] (1.5, 2.5) -- (-4, 0.8);
    \draw[<-,thick] (1.5, -2.5) -- (-4, -0.8);
    \draw[<-, thick] (4.3, 1.8) -- (4.3,-1.8);
    \draw[->, thick] (4.7, 1.8) -- (4.7,-1.8);
    \node[align=left] at (6.5,0) {
    $\delta X=\delta X(\delta E, \delta B),$\\
    $\delta Y=\delta Y(\delta E, \delta B),$\\
    \relax [\eqref{XbyEB}, \eqref{YbyEB}]
    };
    \node[align=left] at (2.5,0) {
    $\delta E=\delta E(\delta X, \delta Y),$\\
    $\delta B=\delta B(\delta X, \delta Y),$\\
    \relax [\eqref{X_EB_first}, \eqref{Y_EB_first},\\
    \relax ~\eqref{X_EB_second}, \eqref{Y_EB_second}]};
    \node[align=left] at (-2.6,2.2) {St\"{u}ckelberg trick for time Diff\\
    \hspace{1cm}
    $\tau \to \tau+ \pi$
    };
    \node[align=left] at (-2,-2.2) {Gauge fixing \\
    $\phi=\bar{\phi}(\tau)\,, ~\delta^{aA}E^a_{[\mu}e^A_{\nu]}=0$
    };
    \end{tikzpicture} 
    \caption{The relations between different forms of Lagrangian.}
    \label{fig:F_EB_XY}
\end{figure}

Here we discuss how the Lagrangian expressed in one basis can be mapped to that in another basis. The relations are summarized in Fig.~\ref{fig:F_EB_XY}.
The arrows in the figure represent mappings between different bases, which we explain one by one in the following.

We first explain the arrow pointing from the left to the bottom right in Fig.~\ref{fig:F_EB_XY}.
Starting with the Lagrangian for $\phi$ and $F^a_{\mu\nu}$, the Lagrangian in the $E$-$B$ basis is obtained by fixing the gauge $\phi=\bar{\phi}(\tau), \delta^{aA}E^a_{[\mu}e^A_{\nu]}=0$ and using the relation
\begin{align}
    F^a_{\mu\nu}&=2n_{[\mu} E^a_{\nu]}+\epsilon_{\mu\nu}{}^{\rho}B^a_{\rho}\,,
    \nonumber \\
    &=e^a_{\alpha}(2n_{[\mu} \delta E_{\nu]}{}^{\alpha}+\epsilon_{\mu\nu}{}^{\rho}\delta B_{\rho}{}^{\alpha}) +\bar{F}^a_{\mu\nu}\,,
\end{align}
where $\epsilon_{\mu\nu\rho}\equiv -n^{\alpha}\epsilon_{\alpha\mu\nu\rho}$ is the spatial Levi-Civita tensor and
\begin{align}
    \bar{F}^a_{\mu\nu} \equiv 2\bar{E}n_{[\mu}e^a_{\nu]}+\bar{B} \epsilon_{\mu\nu}{}^{\rho} e^{a}_{\rho}\,.
\label{Fzero}
\end{align}
The triads are used to trade the gauge index $a$ to the spacetime index $\mu$. The $SU(2)_I$ invariance of the Lagrangian implies that the gauge index must be contracted by another gauge index, which ends up with $e^a_{\mu}e^a_{\nu}=h_{\mu\nu}$. Therefore, the triads do not explicitly appear in the Lagrangian for $\delta E_{\mu\nu}$ and $\delta B_{\mu\nu}$, giving the Lagrangian of the form ${{\cal{L}}_M}(h_{\mu\nu},\tau ,\delta g^{00}, \delta E_{\mu\nu}, \delta B_{\mu\nu})$. 

Next, we explain the mapping from the top right to the left in Fig.~\ref{fig:F_EB_XY}. The broken gauge symmetries can be restored by the St\"{u}ckelberg trick. Although one can apply the St\"{u}ckelberg trick to the $E$-$B$ basis to restore both the time diffeomorphism and the independent rotations of $SU(2)_I$ and $SO(3)_S$, the $SO(3)_S$ invariance is trivial when the theory is written by the metric. Hence, one of the simplest ways would be moving to the $X$-$Y$ basis to manifest the $SU(2)_I$ invariance and then only recovering the time diffeomorphism according to $\tau \to \tilde{\tau}=\tau+\pi$ where
\begin{align}
g^{00} &\to g^{\mu\nu}\nabla_{\mu}\tilde{\tau} \nabla_{\nu}\tilde{\tau} = g^{00}-\frac{2}{a}n^{\mu}\nabla_{\mu}\pi + \cdots
\,, \label{g00pi} \\
h_{\mu\nu} &\to g_{\mu\nu}+\frac{\nabla_{\mu}\tilde{\tau}\nabla_{\nu}\tilde{\tau}}{-(\nabla \tilde{\tau})^2} =h_{\mu\nu}-2a(n_{(\mu}\nabla_{\nu)}\pi + n_{\mu}n_{\nu}n^{\alpha}\nabla_{\alpha}\pi ) +\cdots
\,, \\
    X_{\mu\nu} &\to \frac{\nabla^{\alpha}\tilde{\tau}\nabla^{\beta}\tilde{\tau}}{-2(\nabla \tilde{\tau})^2} (F^a_{\mu\alpha}F^a_{\nu\beta}-\widetilde{F}^a_{\mu\alpha}\widetilde{F}^a_{\nu\beta})=X_{\mu\nu}-\frac{2a}{3}\bar{X}(n_{(\mu}\nabla_{\nu)}\pi+n_{\mu}n_{\nu}n^{\alpha}\nabla_{\alpha}\pi) + \cdots \,, 
    \\
    Y_{\mu\nu} &\to  \frac{\nabla^{\alpha}\tilde{\tau}\nabla^{\beta}\tilde{\tau}}{-(\nabla \tilde{\tau})^2} F^a_{\mu\alpha}\widetilde{F}^a_{\nu\beta}
    =Y_{\mu\nu}-\frac{2a}{3}\bar{Y}(n_{\mu}n_{\nu}n^{\alpha}\nabla_{\alpha}\pi + n_{(\mu}\nabla_{\nu)}\pi) +a (\bar{E}^2+\bar{B}^2)\epsilon_{\mu\nu\rho}\nabla^{\rho}\pi \cdots
    \,.
    \label{Ypi}
\end{align}
Here, $\cdots$ are terms of higher orders in perturbations.

The remaining mapping in Fig.~\ref{fig:F_EB_XY}, between the top right and bottom right, has been already discussed in Section~\ref{Sec_BD_gauge_field}.
To express the Lagrangian in the $X$-$Y$ basis in terms of $E$ and $B$ (i.e., from top right to bottom right in the figure), one can use Eqs.~\eqref{XbyEB} and \eqref{YbyEB}. Conversely, to rewrite the $E$-$B$ basis in terms of $X$ and $Y$, one uses Eqs.~\eqref{X_EB_first}, \eqref{Y_EB_first}, \eqref{X_EB_second}, and \eqref{Y_EB_second}.

Now that we have explained the general procedure for mapping between bases, let us illustrate its application through five concrete examples.
As we will see, all the operators appearing in our EFT action~\eqref{EFT_F2} can be generated by non-linear interactions in the $\phi$–$F$ basis.
\begin{enumerate}
\item Kinetic coupling of gauge fields

The first example is the kinetic term of the $SU(2)$ gauge fields coupled to a scalar field,
\begin{align}
-\frac{1}{4}f_1(\phi)F_{\mu\nu}^aF^{a\mu\nu}&=f_1(\bar{\phi}(\tau)) \left[ \frac{3}{2}(\bar{E}^2-\bar{B}^2) +\bar{E} \delta E -\bar{B} \delta B +\frac{1}{2}(\delta E_{\mu\nu}\delta E^{\mu\nu} -\delta B_{\mu\nu}\delta B^{\mu\nu})\right]\,,
\nonumber 
\\
&
=f_1(\bar{\phi}(\tau)) X\,,
\label{Kinetic_term}
\end{align}
where $\delta E=\delta E^{\mu}{}_{\mu}$ and $\delta B=\delta B^{\mu}{}_{\mu}$. As we mentioned in \eqref{X_FF}, this coupling is exactly mapped into the linear $X$ operator.

\item Chern-Simons coupling

The second example is the Chern-Simons coupling,
\begin{align}
-\frac{1}{4}f_2(\phi)F_{\mu\nu}^a\widetilde{F}^{a\mu\nu}
&=f_2(\bar{\phi}(\tau)) \left[ 3\bar{E}\bar{B} +\bar{E} \delta B + \bar{B} \delta E + \delta E_{\mu\nu}\delta B^{\mu\nu} \right]\,,
\nonumber \\
&=f_2(\bar{\phi}(\tau)) Y
\,,
\label{Chern_Simons}
\end{align}
where these relations are exact according to \eqref{Y_FFtilde}.

\item General function of $F_{\mu\nu}^a F^{a\mu\nu}$ and $F_{\mu\nu}^a\widetilde{F}^{a\mu\nu}$

Thanks to the relations \eqref{X_FF} and \eqref{Y_FFtilde}, any function of $F_{\mu\nu}^a F^{a\mu\nu}$ and $F_{\mu\nu}^a\widetilde{F}^{a\mu\nu}$ can be easily transformed into the $X$-$Y$ basis as
\begin{align}
{\cal{F}}&(-F_{\mu\nu}^a F^{a\mu\nu}/4, -F_{\mu\nu}^a\widetilde{F}^{a\mu\nu}/4)
\nonumber \\
&={\cal{F}}(X,Y)
\notag
\\
&=\bar{\cal{F}}+\bar{\cal{F}}_X\delta X+\bar{\cal{F}}_Y\delta Y
+\frac{1}{2}\bar{\cal{F}}_{XX}\delta X^2+\bar{\cal{F}}_{XY}\delta X\delta Y+\frac{1}{2}\bar{\cal{F}}_{YY}\delta Y^2+\cdots
\notag
\\
&
=\bar{\cal{F}}-\bar{\cal{F}}_X \bar{X}-\bar{\cal{F}}_Y \bar{Y}-\frac{1}{4}\bar{\cal{F}}_X F^a_{\mu\nu}F^{a\mu\nu}-\frac{1}{4}\bar{\cal{F}}_Y F_{\mu\nu}^a\widetilde{F}^{a\mu\nu}
\notag
\\
&
+\frac{1}{2}\bar{\cal{F}}_{XX}\delta X^2+\bar{\cal{F}}_{XY} \delta X\delta Y+\frac{1}{2}\bar{\cal{F}}_{YY} \delta Y^2+\cdots\,,
\label{general_function_of_FF}
\end{align}
where the subscripts denote the derivatives with respect to $X$ and $Y$ and the ellipsis stands for terms of cubic and higher order in perturbations. 
Note that the gauge-flation~\cite{Maleknejad:2011jw,Maleknejad:2011sq} is in this category, where 
\begin{align}
    {\cal{F}}_{\text{gauge-flation}} &=X+\frac{1}{M^4}Y^2
    \nonumber \\
    &=-\frac{1}{M^4}\bar{Y}^2-\frac{1}{4}F^a_{\mu\nu}F^{a\mu\nu} -\frac{\bar{Y}}{2M^4}F_{\mu\nu}^a\widetilde{F}^{a\mu\nu} + \frac{1}{M^4}\delta Y^2
    \,.
    \label{L_gaugeflation}
\end{align}
This also has the Chern-Simons-type tadpole term when expanded around the background even though the Chern-Simons coupling is not present in the covariant form.

\item Other self-interactions

We can also consider higher-order self-interactions other than Eq.~\eqref{general_function_of_FF}. 
We provide two specific examples while omitting detailed calculations.
\begin{itemize}
\item $(F^aF^b)^2$
\ba
&&
\hspace{-0.5cm}
\frac{1}{16}F^{a}_{\mu\nu}F^{b\mu\nu}F^a_{\rho\sigma}F^{b\rho\sigma}=-\frac{1}{3}\bar{X}^2-\frac{1}{6}\bar{X} F^a_{\mu\nu}F^{a\mu\nu}+\delta X_{\mu\nu}\delta X^{\mu\nu}+\cdots
\,.
\label{F4op}
\ea

\item 
$[abc]F^aF^bF^c$
\begin{align}
&
\frac{1}{3!}[abc]F^{a\mu}{}_{\nu}F^{b\nu}{}_{\rho}F^{c\rho}{}_{\mu}
\notag
\\
&
=\frac{1}{3}(\bar{B}\bar{X}+\bar{E}\bar{Y})+\frac{\bar{B}}{4}F^a_{\mu\nu}F^{a\mu\nu}+\frac{\bar{E}}{4}F_{\mu\nu}^a\widetilde{F}^{a\mu\nu}
\notag
\\
&
+\frac{1}{\bar{E}^2+\bar{B}^2}\Biggl[\frac{\bar{B}}{2}(\delta X^2- \delta Y^2)-\bar{E} \delta X \delta Y
\notag
\\
&
\hspace{2cm}
-\bar{B}(\delta X_{\mu\nu}\delta X^{\mu\nu}-\delta Y_{(\mu\nu)}\delta Y^{(\mu\nu)})+2\bar{E}\delta X_{\mu\nu} \delta Y^{\mu\nu} \biggl] +\cdots
\,.
\label{F3op}
\end{align}
Here we used the relation ${
[abc]e^a_\mu e^b_\nu e^c_\rho=\epsilon_{\mu\nu\rho}}$.
\end{itemize}
One can see that new types of quadratic terms $\delta X_{\mu\nu}\delta X^{\mu\nu}$, $\delta X_{\mu\nu}\delta Y^{\mu\nu}$, and $\delta Y_{(\mu\nu)}\delta Y^{(\mu\nu)}$ appear in these cases.

\item Kinetic mixing between scalar and gauge fields 

Finally, we consider kinetic mixing between the scalar field and the gauge fields. A simple example~\cite{Michelotti:2024bbc} is
\begin{align}
&
\frac{1}{2}\nabla_{\mu}\phi \nabla_{\nu}\phi F^{a \mu}{}_{\alpha}F^{a \alpha\nu}
\notag
\\
&= \bar{\phi}'{}^2\left[-\frac{3\bar{E}^2}{2a^2}+\frac{3}{2} \bar{E}^2
\delta g^{00}-\frac{\bar{E}}{a^2}\delta E + \bar{E} \delta g^{00}\delta E - \frac{1}{2a^2} \delta E_{\mu\nu}\delta E^{\mu\nu} +\cdots \right]
\notag
\\
&
= \bar{\phi}'{}^2\Biggl[ {
\frac{3\bar{E}^2}{2a^2}}+\frac{3\bar{E}^2}{2}g^{00} +\frac{\bar{E}^2}{4a^2(\bar{E}^2+\bar{B}^2)}F^a_{\mu\nu}F^{a\mu\nu}+\frac{\bar{E}\bar{B}}{4a^2(\bar{E}^2+\bar{B}^2)}F^a_{\mu\nu}\widetilde{F}^{a\mu\nu}
\notag
\\
&
+ \frac{1}{\bar{E}^2+\bar{B}^2}\left( \bar{E}^2\delta g^{00}\delta X +\bar{E}\bar{B}\delta g^{00}\delta Y -\frac{1}{2a^2} \delta Y_{[\mu\nu]}\delta Y^{[\mu\nu]} \right)
\notag
\\
&
-\frac{2}{9a^2(\bar{E}^2+\bar{B}^2)^3}\left( \bar{Y}^2 \delta X_{\mu\nu}\delta X^{\mu\nu} + \bar{X}^2 \delta Y_{(\mu\nu)}\delta Y^{(\mu\nu)} - 2\bar{X} \bar{Y} \delta X_{\mu\nu} \delta Y^{\mu\nu} \right)+\cdots \Biggl]\,.
\end{align}
This generates new quadratic operators $\delta g^{00}\delta X, \delta g^{00}\delta Y$, and $\delta Y_{[\mu\nu]}\delta Y^{[\mu\nu]}$. Hence, we have seen that all the operators in our EFT action~\eqref{EFT_F2} can be generated by non-linear interactions.

\end{enumerate}

By performing the same procedure, one can rewrite any Lagrangian for $\phi$ and $F^a_{\mu\nu}$ into our EFT language. 
Although we have only focused on the leading order of the derivative expansion, it is possible to extend the analysis into higher orders, such as models including $\delta K_{\mu\nu}\delta X^{\mu\nu}, \delta K_{\mu\nu}\delta Y^{\mu\nu}$ terms, etc.
We leave a more systematic investigation for future work.

\subsection{Clockless theories}
\label{sec:no_clock}
In the case of chromo-natural inflation, there is a scalar field $\phi$, which can be identified with the clock field in our EFT. Even though we have argued that the gauge-flation can be embedded within this framework, one might wonder how this is possible, given that gauge-flation lacks a scalar field.
The key point is that the clock field in the EFT is not necessarily a fundamental scalar field but can be a composite field. One can use  $F_{\mu\nu}^a F^{a\mu\nu}, F_{\mu\nu}^a\widetilde{F}^{a\mu\nu}$ or any other scalar combinations as a clock field as long as it has a background time-dependence. Nonetheless, practically, this type of formulation would not be very useful because the counting of the leading operator becomes obscure. For instance, let us take $Y=-F_{\mu\nu}^a\widetilde{F}^{a\mu\nu}/4$ as a clock field. Then, the normal vector $n_{\mu} \propto \partial_{\mu} Y$ contains second-derivatives of the gauge fields, so the building blocks involve higher derivatives. We thus take an alternative path by following the idea of \cite{Aoki:2021wew}: we first construct the Lagrangian as if a clock field is present and then impose the condition that the Lagrangian is indeed independent of the clock field. This no-clock condition is equivalent to imposing the invariance under the time diffeomorphism. Hence, in this approach, the gauge-flation and its extensions are understood as the EFT with the enlarged symmetry.

To illustrate how this approach works in practice, let us consider a simple example:
\begin{align}
    {\cal{L}}=-\Lambda(\tau)-\frac{1}{4}F^a_{\mu\nu}F^{a\mu\nu}-\frac{\lambda_Y(\tau)}{4}F^{a}_{\mu\nu}\widetilde{F}^{a\mu\nu}+\frac{1}{M^4_{YY}(\tau)}\delta Y^2 + \cdots
    \,,
\end{align}
which include all necessary operators for the gauge-flation \eqref{L_gaugeflation}. We now introduce the St\"{u}ckelberg field
\begin{align}
    \cal{L} &\to -\Lambda(\tau+\pi)-\frac{1}{4}F^a_{\mu\nu}F^{a\mu\nu}-\frac{\lambda_Y(\tau+\pi)}{4}F^{a}_{\mu\nu}\widetilde{F}^{a\mu\nu}+\frac{1}{M^4_{YY}(\tau+\pi)}[Y-\bar{Y}(\tau+\pi)]^2 +\cdots
    \nonumber \\
    &=-\Lambda(\tau)-\frac{1}{4}F^a_{\mu\nu}F^{a\mu\nu}-\frac{\lambda_Y(\tau)}{4}F^{a}_{\mu\nu}\widetilde{F}^{a\mu\nu}+\frac{1}{M^4_{YY}(\tau)}\delta Y^2
    \nonumber \\
    &+ \left[ -\Lambda'(\tau) + \bar{Y}(\tau) \lambda_Y'(\tau) \right]\pi+ \left[\lambda_Y'(\tau)-\frac{2\bar{Y}'(\tau)}{M_{YY}^4(\tau)} \right] \pi \delta Y +\cdots 
    \label{FY_diff}
    \,.
\end{align}
However, apart from the gauge field already included, the theory contains no additional scalar field or composite field that could correspond to $\pi$.
Therefore, the Lagrangian should be independent of the St\"{u}ckelberg field $\pi$.
The condition is given by
\begin{align}
    \Lambda'-\bar{Y}\lambda_Y'=0\,, \qquad 
    \lambda_Y'-\frac{2\bar{Y}'}{M_{YY}^4}  =0
    \,.
    \label{consistency_Y}
\end{align}
On the other hand, using \eqref{general_function_of_FF}, the EFT coefficients for $F=X+\mathcal{F}(Y)$ read
\begin{align}
    \Lambda=-\bar{\mathcal{F}}(\bar{Y})+\bar{Y}\bar{\mathcal{F}}_Y(\bar{Y}) \,, \qquad 
    \lambda_Y=\bar{\mathcal{F}}_Y(\bar{Y}) \,, \qquad \frac{1}{M_{YY}^4}=\frac{1}{2}\bar{\mathcal{F}}_{YY}(\bar{Y}) 
    \,,
\end{align}
by which one can easily confirm that the relations~\eqref{consistency_Y} hold.

Several remarks are in order. First, the ellipsis of \eqref{FY_diff} contains the term proportional to $\pi^2$:
\begin{align}
    \frac{1}{2}\left(\Lambda''+\bar{Y}\lambda_Y'' + \frac{2\bar{Y}'{}^2}{M_{YY}^4} \right) \pi^2
    \,.
\end{align}
This coefficient vanishes when we impose the condition \eqref{consistency_Y}. This is not surprising because it would be sufficient to impose the invariance of the action under infinitesimal transformations: the non-linear terms of $\pi$ will not yield additional conditions. Second, one can find the cubic term $\delta Y^2 \pi$ whose coefficient should vanish to have the time diffeomorphism invariance. However, this coefficient is affected by higher-order operators such as $\delta Y^3$. In other words, the condition to vanish $\delta Y^2 \pi$ yields the condition for higher-order operators, which is not relevant if we are interested in linear perturbations (i.e., the quadratic order action). In this way, we can systematically impose the conditions for no clock field for each order in perturbations.

We conclude this section by studying the no-clock conditions of the general Lagrangian \eqref{EFT_F2}. Using \eqref{g00pi}-\eqref{Ypi}, the transformation laws of the building blocks up to the necessary order are
\begin{align}
    \delta g^{00} &\to \delta g^{00}-\frac{2}{a}n^{\mu}\nabla_{\mu}\pi - \frac{2a'}{a^3}\pi \,, \\
    \delta X_{\mu\nu} & \to \delta X_{\mu\nu}-\frac{\bar{X}'}{3} \pi h_{\mu\nu}
    \,, \\
    \delta Y_{\mu\nu}& \to \delta Y_{\mu\nu}-\frac{\bar{Y}'}{3} \pi h_{\mu\nu} +a(\bar{E}^2+\bar{B}^2) \epsilon_{\mu\nu\rho}\nabla^{\rho}\pi
    \,.
\end{align}
Then, the invariance of \eqref{EFT_F2} under the infinitesimal time diffeomorphisms $\tau \to \tau +\pi$ leads to
\begin{align}
    c&=M_{gg}^4=M_{gX}^2=M_{gY}^2=0
    \,, \quad
    \frac{1}{\hat{M}_{YY}^4}=0\,, \label{no_clockg} \\
    \Lambda' &= \bar{X} \lambda_X' +\bar{Y} \lambda_Y' \,, 
    \label{no_clockL}\\
    \lambda_X' &=2\bar{X}'\left( \frac{1}{M_{XX}^4} + \frac{1}{3\bar{M}_{XX}^4}\right)+\bar{Y}'\left( \frac{1}{M_{XY}^4} + \frac{1}{3\bar{M}_{XY}^4}\right)\,, 
    \label{no_clockX} \\
    \lambda_Y' &=2\bar{Y}'\left( \frac{1}{M_{YY}^4} + \frac{1}{3\bar{M}_{YY}^4}\right)+\bar{X}'\left( \frac{1}{M_{XY}^4} + \frac{1}{3\bar{M}_{XY}^4}\right)
    \,.
    \label{no_clockY}
\end{align}
The operators associated with $\delta g^{00}$ must vanish as anticipated by the fact that the field strength is given by $E$ and $B$, and thus $X$ and $Y$ only. One can also notice that the anti-symmetric operator $\delta Y_{[\mu\nu]}\delta Y^{[\mu\nu]}/M_{YY}^4$ is required to vanish. This is not so apparent but consistent with the results in Sec.~\ref{Sec_mapping} that $\delta Y_{[\mu\nu]}\delta Y^{[\mu\nu]}$ appeared only when considering the kinetic mixing. In addition, one can confirm that the self-interactions \eqref{F4op} and \eqref{F3op} satisfy the no-clock conditions \eqref{no_clockg}-\eqref{no_clockY}, so they are all consistent with Sec.~\ref{Sec_mapping}. It should be also stressed that \eqref{no_clockg}-\eqref{no_clockY} are general, not relying on particular interactions. The generic scalarless (clockless) models ${\cal{L}}={\cal{L}}(g_{\mu\nu},F^a_{\mu\nu})$ are defined to satisfy \eqref{no_clockg}-\eqref{no_clockY} in the framework of the EFT.

\section{Tensor Perturbations and Parity Violation}
\label{Sec_EFT_of_CGW}
In the previous sections, we have constructed the EFT action in the presence of the isotropic configuration of background gauge fields.
This section examines its phenomenological application, especially focusing on tensor perturbations.  
Note that we can safely ignore the quadratic terms except for the third line of \eqref{EFT_F2} because $\delta g^{00},\delta X, \delta Y,$ and $ \delta Y_{[\mu\nu]}$ do not contain tensor perturbations at the linear order. Hence, the EFT action reduces to
\ba
{\cal{S}}=\int{\rd}\tau{\rd}^3 x\sqrt{-g}\Biggl[\frac{\Mpl^2}{2}R-c(\tau)g^{00}-\Lambda(\tau)-\frac{\lambda_X(\tau)}{4} F_{\mu\nu}^aF^{a\mu\nu}-\frac{\lambda_Y(\tau)}{4} F_{\mu\nu}^a\widetilde{F}^{a\mu\nu}
\notag
\\
+\frac{1}{\bar{M}_{XX}^4(\tau)}\delta X_{\mu\nu}\delta X^{\mu\nu}+\frac{1}{\bar{M}_{XY}^4(\tau)}\delta X_{\mu\nu}\delta Y^{\mu\nu}+\frac{1}{\bar{M}_{YY}^4(\tau)}\delta Y_{(\mu\nu)}\delta Y^{(\mu\nu)}+\cdots\Biggr],
\label{EFTofCGW}
\ea
which describes the most general action governing the behavior of gravitational waves in the broken phase at the leading order in the derivative expansion.\footnote{Note that the action with \eqref{EFT_full} and \eqref{EFT_F2} are applicable to other sectors of perturbations. We again note that operators containing more than one derivative of the fields such as $\delta K_{\mu\nu}$ are ignored in the present paper.}

\subsection{Background dynamics}
\label{Backgrounddynamics}
Before addressing tensor perturbations, we briefly study the background dynamics which is determined by only the first line of~\eqref{EFTofCGW}, namely the conditions for the tadpole cancellations. We recall that the background configurations are
\begin{align}
    \bar{g}_{\mu\nu}={\rm diag}[-a^2,a^2,a^2,a^2]\,, \qquad \bar{A}^a_{\mu}=a Q \delta^a_{\mu}
    \,,
    \label{backgrounds}
\end{align}
Then, the tadpole cancellations of \eqref{EFTofCGW} yield
\begin{align}
3\Mpl^2{\cal H}^2&=c+a^2 \Lambda+a^2\left( \rho_E+\rho_B \right)\,,
\label{BG1}
\\
3\Mpl^2({\cal{H}}^2+{\cal{H}}')&=-c+2a^2\Lambda\,,
\label{BG2}
\end{align}
and
\begin{align}
Q''+2{\cal{H}}\left(1-\beta_X\right)Q'+\left[(1-2\beta_X){\cal{H}}^2+{\cal{H}}'\right]Q-2{a}\beta_Y {\cal{H}}{\cal{G}}Q^2+2a^2{\cal{G}}^2Q^3=0\,,
\label{BG3}
\end{align}
where we introduced
\be
\beta_X\equiv -\frac{\lambda_X'}{2\lambda_X{\cal{H}}}\,,
\quad
\beta_Y\equiv -\frac{\lambda_Y'}{2\lambda_X{\cal{H}}}\,,
\quad
\rho_E\equiv \frac{3}{2}\lambda_X\bar{E}^2\,,
\quad
\rho_B\equiv \frac{3}{2}\lambda_X \bar{B}^2\,.
\ee
Here, $\rho_E$ and $\rho_B$ can be understood as the background energy densities of ``electric'' and ``magnetic'' fields, respectively, with $\bar{E}$ and $\bar{B}$ being their amplitudes defined in Eq.~\eqref{EbarBbar}. Note that the EFT parameters $c$ and $\Lambda$ can also be implicit functions of $\bar{E}$ and $\bar{B}$; for example, the gauge-flation yields $\Lambda=-\bar{Y}^2/M^4=-\bar{E}^2\bar{B}^2/M^4$. Nonetheless, it is convenient to use $\rho_E$ and $\rho_B$ as measures of the background gauge fields, as they play a crucial role in the parity violation and the mixing between metric and gauge field perturbations as we will see in the next subsections.

We solve \eqref{BG1} and \eqref{BG2} for $c$ and $\Lambda$ to fix the tadpole EFT coefficients,
\begin{align}
    c&= -\frac{2a^2}{3}(\rho_E+\rho_B) - a^2\Mpl^2 \dot{H}
    \,,\\
    \Lambda&=-\frac{1}{3}(\rho_E+\rho_B) +\Mpl^2(3H^2+\dot{H})\,,
\end{align}
where $H=a'/a^2=\dot{a}/a$ is the Hubble expansion rate and the dot denotes the derivative with respect to the cosmic time, e.g.,~$\dot{a}=a'/a$. 

As we discussed in Sec.~\ref{sec:no_clock}, the clockless theories must satisfy the conditions \eqref{no_clockg}-\eqref{no_clockY}. The relevant conditions for the background and the tensor perturbations are $c=0$ and \eqref{no_clockL}, the former leading to the constraint on the background
\begin{align}
    \Mpl^2\dot{H}=-\frac{2}{3}(\rho_E+\rho_B)\,, \qquad \text{(for $c=0$,\ {\rm e.g.,\ clockless\ theories})}.
    \label{no_clock}
\end{align}
Note that the condition \eqref{no_clockL} automatically holds under the background equations with $c=0$. Hence, the only difference between the generic theories and the clockless theories is whether the background is restricted to satisfy \eqref{no_clock} or not. 

\subsection{Tensor perturbations}
\label{TensorPerturbation}
We next derive the quadratic action for tensor perturbations from the EFT action.
Without loss of generality, we can assume that the tensor perturbations propagate along the $z$-axis. Then the metric can be rewritten as
\be
g_{\mu\nu} = 
\begin{pmatrix}
-a(\tau)^2 & 0 & 0 & 0 \\
0 & a(\tau)^2 [1 + h_+(\tau,z)] & a(\tau)^2 h_\times(\tau,z) & 0 \\
0 & a(\tau)^2 h_\times(\tau,z) & a(\tau)^2 [1 - h_+(\tau,z)] & 0 \\
0 & 0 & 0 & a(\tau)^2
\end{pmatrix}\,,
\ee
where $h_{+}$ and $h_\times$ are the tensor perturbations in the gravity sector, i.e., GWs. Since the gauge index can be identified with the (local) spatial index around the background~\eqref{backgrounds}, the gauge fields also contain the tensor perturbations~\cite{Dimastrogiovanni:2012ew,Adshead:2013qp},
\be
\begin{pmatrix}
A^1_\mu \\
A^2_\mu \\
A^3_\mu
\end{pmatrix}
=
\begin{pmatrix}
0 & a(\tau)[Q(\tau)+t_+(\tau,z)] & a(\tau)t_\times(\tau,z) & 0 \\
0 & a(\tau)t_\times(\tau,z) & a(\tau)[Q(\tau)-t_+(\tau,z)] & 0 \\
0 & 0 & 0 & a(\tau)Q(\tau) 
\end{pmatrix}\,,
\ee
where $t_+$ and $t_\times$ are the tensor perturbations in the gauge sector. For later convenience, we redefine and rescale these quantities as follows,
\be
h_+ = \frac{h_L + h_R}{\sqrt{2}} \;\;,\;\;
h_\times = \frac{h_L - h_R}{i \sqrt{2}} \;\;,\;\;
t_+ = \frac{t_L + t_R}{\sqrt{2}} \;\;,\;\;
t_\times = \frac{t_L-t_R}{ i \sqrt{2}} \;\;,
\ee
\be
H_{L/R}=\frac{a\Mpl}{\sqrt{2}}h_{L/R},\quad
T_{L/R}=\sqrt{2}a\, t_{L/R}
\,,
\ee
where $L$ and $R$ denote left-handed and right-handed circular polarization modes, respectively.

Using the background equations, the quadratic action for the tensor perturbations is computed as, in the momentum space,
\begin{align}
    S^{\rm TT}=\frac{1}{2}\sum_{A=L,R}\int {\rm d}\tau {\rm d}^3k \,\mathcal{L}^{\rm TT}_A\,,
    \label{Stensor}
\end{align}
with
\begin{align}
\mathcal{L}_A^{\rm TT}&=
\left[|H_A'{}|^2-k^2 |H_A|^2 +\frac{a''}{a}|H_A|^2\right]
\nonumber \\
&+{\lambda_X}
\Biggl[ |T_A'{}|^2+\frac{2a\bar{E}}{\Mpl} \left(H^*_A \left( T_A'\pm \alpha \mathcal{K}_A T_A \right)\rm{+c.c.}\right) -\mathcal{K}_A^2 |T_A|^2
\nonumber \\
&\qquad
+a^2\left( \mathcal{G}^2 Q^2 |T_A|^2+\frac{2\bar{E}^2(1-\alpha^2)}{\Mpl^2}|H_A|^2 \right)\pm 2\beta_Y \mathcal{H} \mathcal{K}_A |T_A|^2
\nonumber \\
&\qquad
+\bar{\lambda}_{XX}
\left|  T_A' \mp \alpha \mathcal{K}_A T_A +\frac{a \bar{E}(1+\alpha^2)}{\Mpl} H_A\right|^2+\bar{\lambda}_{YY}\left| \alpha T_A' \pm \mathcal{K}_A  T_A\right|^2 
\nonumber \\
&\qquad+\frac{\bar{\lambda}_{XY}}{2}\left(\left( T_A^{'*} \mp \alpha \mathcal{K}_A T^*_A +\frac{a \bar{E}(1+\alpha^2)}{\Mpl}  H^*_A\right) \left( \alpha T_A' \pm \mathcal{K}_A  T_A\right)\rm{+c.c.}\right)\Biggl]\,,
\label{Ltensor}
\end{align}
where the left (right) handed modes take the upper (lower) sign, we define the polarization-dependent shifted momentum as
\begin{align}
    \mathcal{K}_A\equiv k \mp a \mathcal{G} Q\,,
\end{align}
and
\begin{align}
    \bar{\lambda}_{XX}&=\frac{2\bar{E}^2}{\lambda_X \bar{M}^4_{XX}}\,, \quad 
    \bar{\lambda}_{XY}=\frac{2\bar{E}^2}{\lambda_X \bar{M}^4_{XY}}\,, \quad 
    \bar{\lambda}_{YY}=\frac{2\bar{E}^2}{\lambda_X \bar{M}^4_{YY}}\,, \\
    \alpha &\equiv \frac{\bar{B}}{\bar{E}}=\frac{\mathcal{G}Q}{(\ln a Q)'/a}
    \,,
\end{align}
are dimensionless time-dependent functions. By taking the high-energy limit, one can see that the conditions to avoid ghost and gradient instabilities are given by
\begin{align}
\lambda_X(1+\bar{\lambda}_{XX}+\alpha \bar{\lambda}_{XY}+\alpha^2 \bar{\lambda}_{YY})&>0
\, \quad ({\rm no\ ghost}), \label{no_ghost} \\
\lambda_X(1-\alpha^2 \bar{\lambda}_{XX} + \alpha \bar{\lambda}_{XY}- \bar{\lambda}_{YY})&>0
\, \quad ({\rm no\ grad.\ inst.}).
\label{no_gradient}
\end{align}
The gauge field operators, not only the manifestly parity-violating terms \begin{align*}
    \lambda_Y Y, \quad \frac{1}{\bar{M}_{XY}}\delta X_{\mu\nu}\delta Y^{\mu\nu}\,,
\end{align*}
but also the apparently parity-preserving ones
\begin{align*}
    \lambda_X X, \quad \frac{1}{\bar{M}_{XX}}\delta X_{\mu\nu}\delta X^{\mu\nu}, \quad \frac{1}{\bar{M}_{XY}}\delta Y_{(\mu\nu)}\delta Y^{(\mu\nu)}
    \,,
\end{align*} 
all generate parity violating terms in \eqref{Ltensor} with the opposite signs for the two circular polarization modes.
In the case of the latter (parity-preserving) operators, the polarization dependence is always proportional to the background magnetic field $\bar{B}$, and thus vanishes in the global $SU(2)$ limit $\mathcal{G} \to 0$ (i.e., the triplet of $U(1)$ gauge fields).

Prior to a more rigorous analysis in the quasi-de Sitter limit, we briefly sketch how this action~\eqref{Stensor} leads to the copious production of chiral gravitational waves.
Ignoring the terms with $\bar{\lambda}_{XX}, \bar{\lambda}_{XY}, \bar{\lambda}_{YY}$, we focus on the following terms in proportion to $|T_A|^2$ in Eq.~\eqref{Ltensor} to study the behavior of the gauge field’s tensor modes:
\begin{align}
{\lambda_X}
\Big[ -\mathcal{K}_A^2 +a^2 \mathcal{G}^2 Q^2 
\pm 2\beta_Y \mathcal{H} \mathcal{K}_A \Big]|T_A|^2
= -\left[k^2\mp2(\alpha+\beta_Y)k\mathcal{H}+2\alpha\beta_Y\mathcal{H}^2\right]|\hat{T}_A|^2\,,
\end{align}
where we canonically normalize the field $T_A= \hat{T}_A/\sqrt{|\lambda_X|}$ and assume $Q=$ const$.$ which leads to $a\mathcal{G}Q=\alpha \mathcal{H}$.
The flipped sign of the second term on the right-hand side indicates that $\hat{T}_L$ undergoes tachyonic instability around the horizon crossing ($k\sim \mathcal{H}$) and is significantly amplified for $\alpha,\beta_Y\gtrsim \mathcal{O}(1)$~\cite{Dimastrogiovanni:2012ew,Adshead:2013qp,Dimastrogiovanni:2016fuu}. In contrast, $\hat{T}_R$ remains stable and unamplified, leading to parity violation in the gauge field tensor mode $\hat{T}_A$.

This parity violation in the gauge sector is transmitted to GWs through the mixing,
which is schematically given by
\begin{align*}
\frac{a\sqrt{|\lambda_X|}\bar{E}}{\Mpl} H_A \partial \hat{T}_A
\sim
 \mathcal{H} \sqrt{\frac{|\rho_{E}|}{\rho_{\rm tot}}}  H_A \partial \hat{T}_A
 \,,
\end{align*}
where $\rho_{\rm tot}\equiv 3\Mpl^2H^2$ is the total energy density. Hence, the size of the mixing is determined by how much $\rho_{E,B}$ dominate the universe.
As $\hat T_A$ and $H_A$ couple only to the same chirality components, the parity violation present in $\hat T_A$ is directly transferred to GWs.
We conclude that our EFT \eqref{EFTofCGW} yields a parity violation of the GWs, namely chiral GWs, at a low-energy scale $k\sim \mathcal{H}$ with the size being controlled by the background parameters $\alpha, \rho_{E,B}/\rho_{\rm tot}$ and the EFT parameters such as $\beta_Y$.

\subsection{Quasi-de Sitter limit}

To investigate the behavior of the tensor modes in more detail, we employ the quasi-de Sitter approximation,
\begin{align}
a \simeq \frac{1}{-H_0 \tau} \,, \qquad Q\simeq Q_{\rm sta}={\rm const.}\,,
\label{deSitterBG}
\end{align}
with $H_0$ being a constant. 
We also assume that the following EFT parameters are almost constant
\begin{align}
    \beta_X,\beta_Y,\bar{\lambda}_{XX},\bar{\lambda}_{XY},\bar{\lambda}_{YY} \simeq {\rm const}
    \,.
    \label{static_lambda}
\end{align}
It implies that $\lambda_X$ evolves as $\lambda_X\propto a^{-2\beta_X}$. 

\subsubsection*{Background relations}
We have ${\cal{H}}'\simeq{\cal{H}}^2$ in the quasi-de Sitter phase. The consistency of \eqref{deSitterBG} with the background equations of motion~\eqref{BG1} and \eqref{BG2} lead to
\ba
3c+2a^2\left(\rho_E+\rho_B\right)=3\Mpl^2({\cal{H}}^2-{\cal{H}}') \ll 3\Mpl^2 \mathcal{H}^2 = c +a^2 \Lambda + a^2(\rho_E +\rho_B)
\label{QDS_BG_sol}
\ea
Therefore, unless one considers a fine-tuning, the quasi-de Sitter approximation \eqref{deSitterBG} requires a hierarchy condition
\begin{align}
    \frac{c}{a^2}, \rho_E,\rho_B \ll \rho_{\rm tot} \simeq \Lambda \simeq {\rm const}
    \,.
\end{align}
We stress again that $\Lambda$ can depend on the background gauge fields in concrete models, and hence this condition does not necessarily imply that the gauge fields are subdominant in the energy budget.

Let us next discuss the equation of motion for $Q$, Eq~\eqref{BG3}. In the quasi-de Sitter limit, the equation is approximated by
\begin{align}
Q_{\rm sta} H_0^2 \left( 1-\beta_X -\beta_Y \alpha +\alpha^2 \right) \simeq 0
\label{extremum}
\end{align}
where we have used $\alpha \simeq \mathcal{G}Q_{\rm sta}/H_0$.
Since we have assumed $H_0,Q_{\rm sta}\neq 0$, the expression inside the parentheses should vanish. Note that in the limit $\mathcal{G}\to 0$, we have $\alpha\to 0$ and the EFT parameter is required to satisfy $\beta_X =1$. 

\subsubsection*{Quadratic Lagrangian and stability}
Using the quasi-de Sitter approximations \eqref{deSitterBG} and \eqref{static_lambda}, we can rewrite the Lagrangian~\eqref{Ltensor} as 
\begin{align}
{\cal{L}}^{TT}_A &=  \Delta_A'{}^{\dagger}
\begin{pmatrix}1 & 0 \\ 0 & 1\end{pmatrix}  
\Delta_A' 
-   k^2 \Delta_A{}^{\dagger}
\begin{pmatrix} 1 & 0 \\ 0 & \frac{\gamma_2}{\gamma_1} \end{pmatrix} 
\Delta_A 
\nonumber \\
&+  \Delta_A'{}^{\dagger} K \Delta_A 
-  \Delta_A^\dagger K  \Delta_A' 
\pm k \Delta_A^\dagger P  \Delta_A
-  \Delta_A^\dagger \Omega  \Delta_A  \,,\qquad
\Delta_A = \left( \begin{array}{c} H_A \\ \hat{T}_A \end{array} \right) \,,
\label{tensor-QD}
\end{align}
where $K, P$ and $\Omega$ are $2\times2$ matrices and their components are given by
\begin{align}
    K_{11}&= K_{22}=0  \,, \\
    K_{12}&=-K_{21}=-\frac{\epsilon_E \mathcal{H}}{\sqrt{2}\gamma_1} (1+\alpha^2 +\gamma_1-\alpha \gamma_3) \,, \\
    P_{11} &=0 \,, \\
    P_{12} &= P_{21}= \frac{\sqrt{2}\epsilon_E \mathcal{H}}{\gamma_1} (\alpha \gamma_2 + \gamma_3) \,, \\
    P_{22}&=\frac{2\mathcal{H}}{\gamma_1} (\beta_Y-\alpha\beta_X +\alpha \gamma_2 + \beta_X \gamma_3) \,,\\
    \Omega_{11}&=-2\mathcal{H}^2\left[1+ \frac{\epsilon_E^2}{\gamma_1}(1+\alpha^2+\gamma_1 -\alpha^2 \gamma_2 -2 \alpha \gamma_3) \right]\,, \\
    \Omega_{12}&=\Omega_{21}= \frac{\epsilon_E\mathcal{H}^2}{\sqrt{2}\gamma_1} \left[2\alpha(\alpha \gamma_2 +\gamma_3) + (1-3\beta_X)(1+\alpha^2+\gamma_1-\alpha \gamma_3)  \right]
    \,, \\
    \Omega_{22}&=\frac{\mathcal{H}^2}{\gamma_1}\left[ 2\alpha \beta_Y -2\alpha^2 +(1-\beta_X)(2\alpha^2+\beta_X \gamma_1 -2\alpha \gamma_3) +\alpha(\alpha \gamma_2+\gamma_3) \right]\,.
\end{align}
Here, we defined the canonically normalized field by
\begin{align}
    \hat{T}_A \equiv T_A{\sqrt{\lambda_X\gamma_1}}
\end{align}
and introduced the parameters
\ba
&&
\gamma_1= 1+\bar{\lambda}_{XX}+\alpha\bar{\lambda}_{XY}+\alpha^2\bar{\lambda}_{YY}\,,
\label{gamma1}
\\
&&
\gamma_2= 1-\alpha^2\bar{\lambda}_{XX}+\alpha\bar{\lambda}_{XY}-\bar{\lambda}_{YY}\,,
\label{gamma2}
\\
&&
\gamma_3= \alpha-\alpha\bar{\lambda}_{XX}+\frac{1}{2}(1-\alpha^2)\bar{\lambda}_{XY}+\alpha\bar{\lambda}_{YY}\,,
\label{gamma3}
\ea
and
\begin{align}
    \epsilon_E \equiv \frac{\sqrt{\lambda_X \gamma_1} \bar{E}}{\sqrt{2}\Mpl H_0}=\text{sign}(\bar{E})\sqrt{\frac{\gamma_1 \rho_E}{\rho_{\rm tot}}} \,.
\end{align}
The parameter $\epsilon_E$ 
controls the mixing between the GW and the gauge field. Note that the three parameters $\gamma_i$ can always be solved for $\bar{\lambda}$'s for any real value of $\alpha$. The stability conditions \eqref{no_ghost} and \eqref{no_gradient} are recast as
\begin{align}
    \lambda_X \gamma_1 >0\,\ ({\rm no\ ghost}), \qquad \lambda_X \gamma_2 > 0\,\ ({\rm no\ grad.\ inst.}).
    \label{stability}
\end{align}

\subsubsection*{Chiral gravitational waves}
Parity violation in the tensor perturbations arises from the matrix $P$ in Eq.~\eqref{tensor-QD}, as its overall sign depends on the circular polarization mode, $L$ and $R$.
We examine the conditions under which parity violation does not manifest itself in the tensor perturbations. This corresponds to the case where the matrix $P$ vanishes, i.e., $P = 0$.
For the non-diagonal component to vanish, we impose
\begin{align}
    \alpha \gamma_2 +\gamma_3=0
    \,.
\end{align}
Using this condition and the background equation \eqref{extremum} to eliminate $\gamma_3$ and $\beta_X$, the last condition $P_{22}=0$ reads
\begin{align}
(\alpha -\beta_Y)(1+\alpha^2 + \alpha^2 \gamma_2) =0
\,.
\end{align}
This yields two branches of solutions where the tensor perturbations remain parity-symmetric. 
The first branch $\beta_Y=\alpha$ corresponds to $\beta_X=1$ according to the background equation \eqref{extremum}.  
The second branch requires $\gamma_2<0$. Combined with the stability condition \eqref{stability}, this leads to $\lambda_X < 0$, which in turn implies $\rho_E, \rho_B < 0$. 
Therefore, in order to maintain parity symmetry in the tensor perturbations, one must either set $\beta_X = 1$ or accept negative $\rho_{E}$ and $\rho_B$.
As a side remark, the condition $\rho_E, \rho_B < 0$ results in a violation of the null energy condition, $\dot{H} > 0$, in theories with $c = 0$ (e.g., the clockless case), as evident from Eq.~\eqref{no_clock}.

Since $P_{11}=0$, parity violation originates in the gauge field tensor mode $\hat{T}_A$ and is subsequently transmitted to GWs through their mixing. Hence, even if $\hat{T}_A$ breaks parity, the GWs may remain parity-symmetric if their mixing can be eliminated.
The mixing arises from the non-diagonal components of $K,P,$ and $\Omega$, which are suppressed by $\epsilon_E$. 
All the non-diagonal components of $K,P,\Omega$ vanish when
\begin{align}
    \alpha \gamma_2 +\gamma_3=0\,,
    \label{no_int1}
\end{align}
and
\begin{align}
1+\alpha^2 +\gamma_1 -\alpha \gamma_3 = 1+\alpha^2 +\gamma_1 +\alpha^2 \gamma_2
=0\,,
\label{no_int2}
\end{align}
where the first condition \eqref{no_int1} has been used for the first equality of \eqref{no_int2}. These conditions cannot be satisfied for $\lambda_X >0$ due to the stability conditions \eqref{stability}. Hence, the GWs and gauge fields can be decoupled only when $\rho_E, \rho_B<0$. 
We also notice that $\epsilon_E$ dependent term of $\Omega_{11}$ vanishes when \eqref{no_int1} and \eqref{no_int2} hold. There is thus no correction to the linear dynamics of GWs in the decouple case.

\subsubsection*{Superhorizon dynamics}
We turn our attention to the superhorizon dynamics of the GWs and gauge fields $(k \ll \mathcal{H})$. The superhorizon equations of motion are
\begin{align}
    H_A'' - 2 \mathcal{H}^2 H_A &= \mathcal{O}(\epsilon_E)
    \,, \\
    \hat{T}_A'' + \Omega_{22}\hat{T}_A&=\mathcal{O}(\epsilon_E)
    \,,
\end{align}
where using the background equation \eqref{extremum}, $\Omega$ is simplified to be
\begin{align}
\Omega_{22}=\frac{\mathcal{H}^2}{\gamma_1}\left[ (1-\beta_X)(2+2\alpha^2+\beta_X \gamma_1 -2\alpha \gamma_3) +\alpha(\alpha \gamma_2+\gamma_3) \right]
\,.
\end{align}
Since the equation of motion for $H_A$ takes the standard form, the tensor perturbations $h_A \propto H_A/a$ are approximately conserved on the super-horizon scales up to the possible $\mathcal{O}(\epsilon_E)$ correction.
In contrast, by neglecting $\epsilon_E$, the solutions of the gauge fields are given by
\begin{align}
    \hat{T}_A \propto a^{-\frac{1}{2}\pm\frac{1}{2}\sqrt{1-4\Omega_{22}/\mathcal{H}^2}}
    \,,
\end{align}
and thus
\begin{align}
t_A \propto \frac{\hat{T}_A}{a\sqrt{|\lambda_X|}} \propto  a^{-\frac{3}{2}+\beta_X\pm \frac{1}{2}\sqrt{1-4\Omega_{22}/\mathcal{H}^2}}
\,.
\end{align}
We see that the first branch of the parity-symmetric condition $\alpha \gamma_2+\gamma_3=0$ and $\alpha-\beta_Y=0=\beta_X-1$ gives
\begin{align}
    t_A \propto a^0\ \ {\rm and}\ \ a^{-2}\,.
\end{align}
The gauge field perturbations in this case are constant or decaying in the super-horizon limit, with only the constant mode persisting at late times.\footnote{In the context of the anisotropic inflation model with a $U(1)$ gauge field, it is known that when $\lambda_X \propto a^{-2}$, both the background electric field and its superhorizon perturbations remain constant~\cite{Watanabe:2009ct}. Our first branch may suggest a non-Abelian analogue, in which a similar mechanism operates for the tensor modes of the gauge field. (However, straightforward non-Abelian extensions of the model do not have the isotropic background~\cite{Murata:2011wv,Gao:2021qwl}.)}
Conversely, whenever the tensor perturbations of the gauge field deviate from the above superhorizon scaling, parity symmetry must be broken and chiral gravitational waves are produced, provided that the effective energy densities are positive ($\rho_E, \rho_B>0$).

In summary, we have found that parity symmetry in the tensor sector can be preserved only under specific conditions: either the gauge kinetic function must be fine-tuned to scale as $\lambda_X\propto a^{-2}$, or the effective energy densities must be negative, i.e., $\rho_E, \rho_B < 0$. The absence of mixing between gravitational waves and gauge fields also requires $\rho_E, \rho_B < 0$.
Taking the contraposition of these statements, we conclude that the symmetry-breaking pattern $SU(2)_I \times SO(3)_S \to SO(3)_D$ inevitably leads to chiral GWs, 
provided that the effective energy densities are positive ($\rho_E, \rho_B > 0$) and the gauge kinetic function does not scale as $\lambda_X\propto a^{-2}$.

\section{Conclusion}
\label{Conclusion}
In this work, we have developed an effective field theory (EFT) of chiral gravitational waves in the presence of (non-)Abelian gauge fields with an isotropic background configuration~\eqref{nontrivial_BGsol} during inflation. The physical VEV~\eqref{nontrivial_BGsol_EB}, composed of ``electric'' and ``magnetic'' fields, spontaneously breaks the gauge symmetry $SU(2)_I$ and the spatial rotation part of local Lorentz symmetry $SO(3)_S$ down to the diagonal subgroup $SO(3)_D$. 
We introduced two equivalent sets of building blocks that respect the residual symmetry: the $E$-$B$ basis and the $X$-$Y$ basis.
While the $X$-$Y$ basis, defined in Eqs.~\eqref{deltaX} and \eqref{deltaY}, is convenient for constructing the EFT Lagrangian~\eqref{EFT_F2}, the $E$-$B$ basis~\eqref{EB_perturbations} is also useful in translating a covariant Lagrangian written in the $\phi$-$F$ basis into the EFT language. The relations among these different bases are summarized in Fig.~\ref{fig:F_EB_XY}.
The resulting EFT~\eqref{EFT_F2} provides the most general quadratic action for perturbations around our background, including chrome-natural inflation~\cite{Adshead:2012kp}, gauge-flation~\cite{Maleknejad:2011jw}, inflation with a triplet of $U(1)$ gauge fields~\cite{Yamamoto:2012tq, Piazza:2017bsd, Firouzjahi:2018wlp} as particular examples, and is expected to serve as a solid foundation for future phenomenological studies of chiral GW production and other observational signatures.

In Sec.\ref{Sec_EFT_of_CGW}, we analyzed tensor perturbations based on the quadratic EFT action for the tensor modes, Eq.~\eqref{Stensor}. Within this framework, we explicitly derived the conditions under which tensor perturbations are free from ghost and gradient instabilities.
A key result of our analysis is that the produced GWs are generically chiral. Nevertheless, we identified two exceptional conditions under which parity symmetry in the tensor sector can be preserved: (i) the gauge kinetic function $\lambda_X$ is negative, which implies a negative effective energy density of the gauge fields $\rho_E, \rho_B<0$, or (ii) $\lambda_X \propto a^{-2}$, in which case the gauge field tensor modes $t_A$ freeze on super-horizon scales.
While we did not estimate the amplitude of chiral GWs in this work, our formalism may accommodate both enhancement and suppression mechanisms of the tensor modes. For instance, tachyonic instability could be made stronger to amplify GWs, or the effective mass of the gauge field tensor modes $\hat{T}_A$ could be increased to reduce GW production. Furthermore, the formalism can be extended to include cubic and quartic interactions in the EFT action, enabling the computation of non-Gaussianities~\cite{Agrawal:2017awz,Dimastrogiovanni:2018xnn,Fujita:2021flu}. We leave such phenomenological applications for future work.

We have not addressed the scalar and vector perturbations in this work. Studying their stability and observables is expected to impose additional constraints on the allowed background dynamics and EFT parameter space, and to provide important phenomenological insights. For instance, in chromo-natural type models, it is known that enhancing chiral GWs at CMB scales often leads to tension with scalar perturbations~\cite{Papageorgiou:2018rfx, Papageorgiou:2019ecb}. Our EFT framework offers a general and systematic way to revisit this issue beyond model-specific analyses.
Moreover, we have limited our construction to the leading order in the derivative expansion, but higher-order terms—such as $\delta K_{\mu\nu} \delta K^{\mu\nu}$ or $\delta K_{\mu\nu} \delta X^{\mu\nu}$— can arise in models that incorporate modified gravity~\cite{Dimastrogiovanni:2023oid,Murata:2024urv}, gauge-gravity couplings, and higher derivative terms, etc.
It would be interesting to explore how such operators affect the properties of chiral GWs. In addition, recent discussions on the breakdown of perturbative treatment~\cite{Dimastrogiovanni:2024lzj} or the emergence of new attractor solutions in the presence of strong backreaction~\cite{Iarygina:2023mtj, Dimastrogiovanni:2025snj} may also benefit from the EFT perspective developed here.
These directions represent promising avenues for future research.

\acknowledgments
We thank Kohei Kamada, Shinji Mukohyama, Tomoaki Murata, Atsuhisa Ota, and Shinji Tsujikawa for useful discussions and comments. 
R.K. thanks the hospitality of the Yukawa Institute for Theoretical Physics (YITP), Kyoto University, where this work was initiated and the Physics Laboratory of the \'{E}cole Normale Sup\'{e}rieure (LPENS) where it was completed.
This work is supported by JSPS KAKENHI grants 
Nos.~20H05852 (KA), 23K03424 (TF), 24KJ2108 (RK), and 24K17046 (KA).

\renewcommand{\theequation}{A.\arabic{equation}}
\setcounter{equation}{0}

\appendix

\section{Symmetry Breaking and Gauge Fixing}
\label{AppendixA}
In this appendix, we explicitly show the spontaneous symmetry breaking and gauge fixing.

To see the remaining and broken parts of global symmetry, let us define a VEV with the electric fields;
\be
\langle E^{aA}\rangle=\langle E_\mu^a e^A_\nu h^{\mu\nu}\rangle=\bar{E}\delta^{aA}\,.
\label{App_VEV}
\ee
Under infinitesimal transformations of independent (global) $SU(2)_I$ and $SO(3)_S$, the VEV is transformed as
\be
\langle E^{aA}\rangle\rightarrow \langle E^{aA}\rangle+[abc]\theta^b\langle E^{cA}\rangle+[ABC]\tilde{\theta}^B\langle E^{aC}\rangle=\langle E^{aA}\rangle +[abA](\theta^b-\tilde{\theta}^b)\bar{E}
\ee
where $\theta^a$ and $\tilde{\theta}^A$ denote the rotation angles of $SU(2)_I$ and $SO(3)_D$, respectively.
From the local isomorphism of $SU(2)$ and $SO(3)$ groups, these two have the same structures in infinitesimal transformations.
Redefining the rotation angles with $\phi^a=\theta^a+\tilde{\theta}^a$ and $\tilde{\phi}^a=\theta^a-\tilde{\theta}^a$, which denote the same and opposite directional rotations of $SU(2)_I$ and $SO(3)_S$, respectively, we can see that the VEV~\eqref{App_VEV} is invariant under $SO(3)_D$ transformation with $\phi^a$ while is not invariant under $SO(3)_A$ transformation with $\tilde{\phi}^a$.
Therefore, $SO(3)_D$ is the remaining symmetry and $SO(3)_A$ is the spontaneously broken symmetry in the VEV.

Next, let us show the gauge fixing~\eqref{gauge_fixing}.
Under infinitesimal transformations of independent (local) $SO(3)_D$ and $SO(3)_A$, $\delta_{aA}E^a_\mu e^A_\nu$ is transformed as
\be
\delta_{aA}E^a_\mu e^A_\nu\rightarrow\delta_{aA}E^a_\mu e^A_\nu-[abc]\tilde{\phi}^b E^{[a}_\mu e^{c]}_\nu=\bar{E}h_{\mu\nu}+\delta_{aA}\delta E^a_\mu e^a_\nu-\bar{E}[abc]\tilde{\phi}^c e^a_\mu e^b_\nu +\cdots\,,
\ee
where dots denote higher order in perturbations.
As we can see, $\delta_{aA}E^a_\mu e^A_\nu$ is invariant quantity under $SO(3)_D$ transformation.
Furthermore, the symmetric part, $\delta_{aA}E^a_{(\mu} e^A_{\nu)}$, is invariant even for $SO(3)_A$ transformation.
By choosing the rotation angle as 
\be
\tilde{\phi}^{a}=\frac{1}{2\Bar{E}}h^{\mu\nu}[abc]e_\mu^b\delta E^c_\nu\,,
\ee
we can take the gauge fixing~\eqref{gauge_fixing}, $\delta_{aA} E^a_{[\mu} e^A_{\nu]}=0$.
Note that this gauge fixing is allowed only when the background electric field $\bar{E}$ takes a non-zero value.
This gauge choice only fixes the local $SO(3)_A$ symmetry.
Note that the broken symmetry, $SO(3)_A$, can be restored by performing the St\"{u}ckelberg trick by promoting the rotation angle $\tilde{\phi}^a$ to Goldstone bosons $\tilde{\pi}^a$.

\bibliographystyle{JHEP}
\bibliography{bib}

\end{document}